\documentclass[aps,pre,reprint,groupedaddress,showpacs,amsmath,amssymb]
{revtex4-1}

\usepackage{graphicx}% Include figure files
\usepackage{bm}% bold math
\newtheorem{remark}{Remark}

\renewcommand{\vec}[1]{\bm{#1}}
\newcommand{\pder}[2]{\frac{\partial #1}{\partial #2}}
\newcommand{\der}[2]{\frac{d #1}{d #2}}
\newcommand{\Gr}{\textit{Gr}}
\newcommand{\Ce}{C_\varepsilon}
\newcommand{\Z}{\mathbb{Z}}
\newcommand{\R}{\mathbb{R}}

\newcommand{\C}{\mathbb{C}}
\newcommand{\stav}[1]{\left\langle #1 \right\rangle}
\newcommand{\htav}[1]{\overline{#1}}
\newcommand{\norm}[1]{\left\| #1 \right\|}
\newcommand{\abs}[1]{\left\vert #1 \right\vert}
\newcommand{\gvec}[1]{\bm{#1}}
\newcommand{\leg}{\mathcal{L}}
\newcommand{\e}{\text{e}}
\renewcommand{\Re}{\text{\rm Re}}
\renewcommand{\Im}{\text{\rm Im}}

\begin{document}

%\preprint{}

\title{Optimal bounds with semidefinite programming:
\\an application to stress driven shear flows}


\author{G. Fantuzzi}
\email[]{gf910@ic.ac.uk}
%\thanks{}
\author{A. Wynn}
\email[]{a.wynn@imperial.ac.uk}
\affiliation{Department of Aeronautics, Imperial College London, 
South Kensington Campus, London, SW7 2AZ, United Kingdom}


\date{\today}

\begin{abstract}

We introduce an innovative numerical technique based on convex optimization
to solve a range of infinite dimensional variational problems arising from 
the application of the background method to fluid flows. In contrast to most
 existing schemes, we do not consider the Euler--Lagrange equations for the 
minimizer. Instead, we use series expansions to formulate a finite dimensional
 semidefinite program (SDP) whose solution converges to that of the original 
variational problem.
%
%The formulation is rigorous, meaning that a solution of the SDP gives
% a certifiably feasible solution for the infinite dimensional problem. 
Our formulation accounts for the influence of all modes in the 
expansion, and the feasible set of the SDP corresponds to a subset of the 
feasible set of the original problem.
Moreover, SDPs can be easily formulated when the fluid is subject to
 imposed boundary fluxes, which pose a challenge for the traditional methods.
We apply this technique to compute rigorous and near-optimal upper bounds on
 the dissipation coefficient for flows driven by a surface stress. We improve
 previous analytical bounds by more than 10 times, and show that the bounds
 become independent of the domain aspect ratio in the limit of vanishing
 viscosity. We also confirm that the dissipation properties of stress driven
 flows are similar to those of flows subject to a body force localized in a
 narrow layer near the surface. Finally, we show that SDP relaxations are an
 efficient method to investigate the energy stability of laminar flows driven
 by a surface stress.
\end{abstract}

%\keywords{}

\maketitle

\section{Introduction \label{S:Introduction}}

Turbulent flows typically exhibit enhanced transport, mixing and/or dissipation
 compared to steady flows, but an accurate characterization of their properties
 is challenging due to their dynamic complexity.
Given the computational cost of full numerical simulations in highly turbulent
 regimes and the absence of closed-form solutions to the Navier--Stokes
 equations, a common approach is to derive rigorous bounds for key turbulent
 properties (e.g. dissipation, transport, etc.) as a function of the forcing
 parameters (e.g. Reynolds number, boundary conditions, body forces, etc.).

Among other techniques, the \emph{background method}~\cite{Constantin1995} has
 been applied to derive rigorous scaling laws directly from the governing
 equations in a wide range of contexts. 
Typical examples include computing bounds for the energy dissipation in shear
 flows~\cite{Doering1992,Doering1994,Nicodemus1998a,Nicodemus1998b,
Nicodemus1998} and for the net turbulent heat transport in Rayleigh-B\'{e}nard
 convection (e.g.~\cite{Doering1996,Whitehead2011}). 
In the context of shear flows, the method relies on the decomposition of the
 flow velocity into a steady \emph{background field} $\phi$, that absorbs any
 inhomogeneous boundary conditions (BCs), plus an arbitrary perturbation
  $\vec{u}$. The bounds (upper or lower) are then expressed in terms of a
 functional $\mathcal{B}\{\phi\}$, and the optimal ones are obtained by
 extremizing this functional subject to the positivity of a $\phi$--dependent
 quadratic form $\mathcal{Q}\{\vec{u}\}$ --- a condition known as the
 \emph{spectral constraint}.

This infinite dimensional variational problem has generally been studied by
 considering the Euler--Lagrange equations for the optimal background field.
 Solving such equations typically requires delicate computations in order to
 avoid spurious solutions that extremize the bound but do not satisfy the
 spectral constraint (see~\cite{Wen2013,Wen2015a} for a detailed discussion).
Recently, a two-step time-marching algorithm has been shown to give the 
correct solution of the Euler--Lagrange equations for a number of canonical 
 examples~\cite{Wen2013,Wen2015a}.

A particular challenge in these computations comes from flows 
subject to imposed boundary fluxes
 (Neumann BCs) or mixed Dirichlet--Neumann BCs. In many such cases the 
 functional $\mathcal{B}$ depends on unknown boundary values of the
 background field (e.g.~\cite{Hagstrom2010,Hagstrom2014,Wittenberg2010a}),
 and the solution of the Euler--Lagrange equations is further complicated by
 the need to enforce so-called
 \emph{natural boundary conditions}~\cite{Courant1953,Giaquinta2004}. 
 The technical difficulties posed by these additional conditions  have not
 yet been addressed, and to our knowledge a fully optimal solution of such
 bounding problems has never been obtained.

In this work, we propose a novel approach to compute near-optimal
 bounds that can be easily applied to flows with fixed boundary fluxes. 
 Our method differs from previous computational techniques because
 it does not consider the Euler--Lagrange equations, so that the complications
 arising from any natural BCs can be avoided. Instead, we develop the approach
 proposed by the authors in~\cite{Fantuzzi2015} and consider the quadratic
 form $\mathcal{Q}$ directly to show that the variational problem can be
 rigorously formulated as a semidefinite program (SDP). Our bounds are
 near-optimal, i.e. they are obtained with a mild restriction on the background
 fields, but are expected to converge to the fully optimal bounds 
 (although we do not provide a formal proof). 
 Moreover, our formulation is rigorous, meaning that 
 the feasible set of the SDP corresponds to
  a subset of the set of background fields that satisfy the infinite dimensional 
 spectral constraint. This means that mathematically rigorous, near-optimal 
 bounds could be obtained by controlling the numerical round-off errors when 
solving the SDP; however, this is outside the scope of the present work. 
Finally, we show that our techniques can also be applied to compute 
energy stability boundaries for laminar flows.

We illustrate our method by computing upper bounds on the dissipation
 coefficient $\Ce$ for two and three dimensional shear flows driven by
 a surface stress. Flows of this kind arise, for example, in physical
 oceanography, when wind blows over a body of water. The three-dimensional
 flow was first studied by Tang \textit{et al.}~\cite{Tang2004}, who used the
 background method to estimate $\Ce \leq {\Gr}\,{(7.531\Gr^{0.5} - 20.3)^{-2}}$
 for large $\Gr$, where the Grashoff number $\Gr$ represents the nondimensional
 forcing. Strictly speaking, however, these bounds apply to a different flow,
 where the imposed stress is approximated by a body force localized near the
 boundary. A bounding problem that incorporates the fixed-shear boundary
 condition was subsequently formulated by 
 Hagstrom\,\&\,Doering~\cite{Hagstrom2014}, who used a piecewise linear
 background field to prove $\Ce \leq 1/16$ for $\Gr\geq 16$ in two dimensions,
 and $\Ce\leq 1/{(2\sqrt{2})}$ uniformly in $\Gr$ for three dimensional flows.
 In this work, we close the circle of ideas and compute near-optimal bounds by
 solving the bounding problem formulated by Hagstrom\,\&\,Doering over a mildly
 restricted (but not piecewise linear) set of background fields.

The rest of this paper is organized as follows. 
 In Section~\ref{S:ShearFlowIntro} we describe the flow and summarize the
 bounding problem derived in~\cite{Hagstrom2014}. 
 Section~\ref{S:ComputationalStrategy} gives a brief overview of semidefinite
 programming and of our computational strategy. We formulate and solve an SDP
 to compute bounds for the two dimensional flow in Section~\ref{S:2DBounds},
 and extend the analysis to the three dimensional case in 
 Section~\ref{S:3DBounds}. In Section~\ref{S:EnergyStability} we illustrate how
 SDPs can be used in energy stability theory by computing the critical $\Gr$
 for the stability of a stress driven Couette flow. 
 Finally, Section~\ref{S:Conclusions} offers concluding remarks.

%%%%%%%%%%%%%%%%%%%%%%%% 
%    EQUATIONS OF MOTION
%%%%%%%%%%%%%%%%%%%%%%%%
\section{Stress-driven shear flow: equations 
\& a bounding principle \label{S:ShearFlowIntro}}

We begin by describing the flow and the bounding principle for the dissipation
 coefficient derived by Hagstrom\,\&\,Doering~\cite{Hagstrom2014}. Dimensional
 quantities will be denoted by the suffix $\star$. We will write all equations
 in three dimensions; the two dimensional case is obtained by simply removing
 any terms related to the $y$ direction.

\subsection{Fluid equations \label{S:FluidEquations}}

We consider an incompressible layer of fluid of constant depth $h$, kinematic
 viscosity $\nu$ and density $\rho$ driven at $z_\star=h$ by a shear stress
 $\tau$ in the $x_\star$ direction. We impose no slip conditions at $z_\star=0$
 and horizontal periodicity across 
 $0\leq x_\star \leq h\Gamma_x$ and $0 \leq y_\star\leq h\Gamma_y$,
 where $\Gamma_x$ and $\Gamma_y$ are the domain aspect ratios. 

Following Tang \textit{et al.}~\cite{Tang2004}, we consider the nondimensional
 variables
\begin{equation}
\vec{x} = \frac{\vec{x}_\star}{h}, 
 \qquad t = \frac{t_\star\nu}{h^2},
 \qquad \vec{u}=\frac{\vec{u}_\star h}{\nu}, 
 \qquad p = \frac{p_\star h^2}{\rho \nu^2}
\end{equation}
and write the Navier--Stokes equations as
\begin{subequations}
\begin{gather}
\label{E:NS}
\pder{\vec{u}}{t} + \vec{u}\cdot \nabla \vec{u} + \nabla p 
= \nabla^2 \vec{u},
\\
\nabla\cdot \vec{u}=0.
\end{gather}
\end{subequations}
The nondimensional velocity 
$\vec{u}\equiv u\vec{\hat{i}}+v\vec{\hat{j}}+w\vec{\hat{k}}$ has period
 $\Gamma_x$ and $\Gamma_y$ in the $x$ and $y$ directions, respectively, and
 satisfies the additional boundary conditions
\begin{equation}
\label{E:BC}
\begin{gathered}
\vec{u}\vert_{z=0} = 0, \,\, \pder{u}{z}\Big\vert_{z=1}= \Gr, \,\,
 \pder{v}{z}\Big\vert_{z=1}=0, \,\, w\vert_{z=1}=0 
\end{gathered}
\end{equation} 
at the top and bottom surfaces, where the Grashoff number $\Gr$ is the
 characteristic nondimensional control parameter of the flow and is defined as
\begin{equation}
\Gr := \frac{\tau h^2}{\rho \nu^2}.
\end{equation} 

The laminar solution to these equations corresponds to the Couette flow
 $\vec{u}_L=\Gr \,z\, \vec{\hat{i}}$. Introducing the horizontal-time and
 space-time averages
\begin{subequations}
\begin{gather}
\htav{q(\vec{x},t)} := 
\lim_{T\to\infty} \frac{1}{T\,\Gamma_x\,\Gamma_y} \int\limits_{0}^{T}
 \int\limits_{0}^{\Gamma_y} \int\limits_{0}^{\Gamma_x} q(\vec{x},t)
 \,dx\,dy\,dt,
\\
\stav{q(\vec{x},t)} := 
\int\limits_{0}^{1} \htav{q}(z) \, dz,
\end{gather}
\end{subequations}
energy stability analysis shows that $\vec{u}_L$ is stable if 
\begin{equation}
\label{E:EnStabConstr}
\stav{ \norm{\nabla\vec{u}}^2 + \Gr\, u w } \geq 0
\end{equation}
for all time-independent, incompressible fields $\vec{u}(x,y,z)$ satisfying the
 homogeneous BCs
\begin{equation}
\label{E:HomogBCs}
\vec{u}\vert_{z=0} = \pder{u}{z}\Big\vert_{z=1} = 
\pder{v}{z}\Big\vert_{z=1}= w\vert_{z=1}=0.
\end{equation} 
Hagstrom\,\&\,Doering~\cite{Hagstrom2014} showed that the Couette profile is
 stable for $\Gr\leq 139.54$ and $\Gr \leq 51.73$ for the two and
 three dimensional cases, respectively.

We describe the flow by the bulk energy dissipation rate per unit mass
\begin{equation}
\varepsilon:=\langle\nu\norm{\nabla_\star\vec{u}_\star}^2\rangle = 
\frac{\nu^3}{h^4}\langle\norm{\nabla\vec{u}}^2\rangle,
\end{equation}
where $\nabla_\star$ is the dimensional gradient, and the associated
 nondimensional dissipation coefficient $\Ce$, defined as
\begin{equation}
\label{E:SkinFrictionDef}
\Ce := \frac{\varepsilon h}{\htav{u_\star}(h)^3} = 
\frac{\Gr}{\htav{u}(1)^2}.
\end{equation}
The last equality follows from the identity 
 $\langle\norm{\nabla\vec{u}}^2 \rangle = \Gr\,\htav{u}(1)$,
 which can be proven by space-time averaging the dot product of the momentum
 equation with $\vec{u}$~\cite{Hagstrom2014}.

\subsection{A bounding problem for $\Ce$ \label{SS:BoundingProblem}}

Tang \textit{et al.}~\cite{Tang2004} demonstrated that the laminar dissipation
 rate $\varepsilon_L = Gr^2\nu^3h^{-4}$ provides a rigorous upper bound on
 $\varepsilon$, corresponding to the lower bound on the dissipation coefficient
 $\Ce \geq 1/Gr$. 
 Note that this bound is valid for both two and three dimensional flows and is
 sharp, being saturated by the Couette profile.

A rigorous upper bound on $\Ce$ was derived by 
 Hagstrom\,\&\,Doering~\cite{Hagstrom2014}, who applied the background method
 to derive a lower bound on $\htav{u}(1)$. 
Specifically, they showed that if a background field $\phi(z)$ can be
 chosen such that
\begin{equation}
\label{E:BPBC}
\phi(0)=0, \qquad \der{\phi}{z}\Big\vert_{1} = \Gr,
\end{equation}
and such that the spectral constraint
\begin{equation}
\label{E:SpectralConstraint}
\mathcal{Q}\{\vec{u},\phi\} := \stav{ \norm{\nabla\vec{u}}^2 
+ 2\der{\phi}{z} u w } \geq 0
\end{equation}
holds for all time-independent, incompressible fields $\vec{u}(x,y,z)$
 satisfying the BCs in~\eqref{E:HomogBCs}, then
\begin{equation}
\htav{u}(1) \geq 2\phi(1) - \frac{1}{\Gr} \int\limits_{0}^{1}\left( 
\der{\phi}{z}\right)^2 dz =: - \mathcal{B}\{\phi\}.
\end{equation}
We say that $\phi$ is a \textit{feasible} background field
 if~\eqref{E:SpectralConstraint} holds for all admissible $\vec{u}$, and that
 $\phi$ is \textit{strictly feasible} if~\eqref{E:SpectralConstraint} holds
 with strict inequality for all admissible $\vec{u}\neq 0$. 

The optimal upper bound on $\Ce$ achievable via the background method then
 follows from~\eqref{E:SkinFrictionDef} after minimizing $\mathcal{B}$ over all
 feasible background fields satisfying~\eqref{E:BPBC}.

\subsection{A rescaled bounding problem in Fourier space 
\label{SS:SCFourier}}

In order to construct a feasible and near-optimal background field numerically,
 it is convenient to rescale the vertical domain to the interval $[-1,1]$ by
 letting $\zeta=2z-1$.  Moreover, the horizontal periodicity allows us to write
 $\vec{u}$ as the Fourier series
\begin{equation}
\vec{u}(x,y,\zeta) = \sum_{m,n\in\Z} \vec{U}_{mn}(\zeta) \, 
\e^{i(\alpha_m x + \beta_n y)}
\end{equation}
where 
$$
\alpha_m = \frac{2\pi m}{\Gamma_x}, \qquad \beta_n = 
\frac{2\pi n}{\Gamma_y}.
$$
For all $m,n\in\Z$ the complex vector-valued functions 
 $\vec{U}_{mn} = U_{mn}\vec{\hat{i}} + V_{mn}\vec{\hat{j}} 
  + W_{mn}\vec{\hat{k}}$ 
 must satisfy the incompressibility condition
\begin{equation}
\label{E:FourierIncompressibility}
i\alpha_m U_{mn} + i\beta_n V_{mn} + 2\der{W_{mn}}{\zeta} = 0
\end{equation}
and the BCs
\begin{equation}
\label{E:FourierBCs}
\vec{U}_{mn}(-1) = \der{U_{mn}}{\zeta}\Big\vert_{1} = 
\der{V_{mn}}{\zeta}\Big\vert_{1}= W_{mn}(1)=0.
\end{equation}
Moreover, the requirements that the Fourier modes combine into a real-valued
 field $\vec{u}$ implies that $\vec{U}_{-m,-n}=\vec{U}_{mn}^*$, 
 where $^*$ denotes complex conjugation.

In addition, we introduce the rescaled and Fourier-transformed 
 gradient-type operator
\begin{equation}
\mathcal{D} := \begin{pmatrix} i\alpha_m, & i\beta_n, & 
\displaystyle 2\der{}{\zeta} \end{pmatrix}
\end{equation}
and define the quadratic forms $\mathcal{Q}_{mn}\{\vec{U}_{mn}, \phi\}$ as
\begin{equation}
\mathcal{Q}_{mn} %\{\vec{U}_{mn}, \phi\}
:=  \int\limits_{-1}^{1} \left[ \norm{\mathcal{D}\vec{U}_{mn}}^2 
+ 4\der{\phi}{\zeta} \Re\left(U_{mn} W_{mn}^* \right) \right] d\zeta
\end{equation}
so that we may write
\begin{align}
\mathcal{Q}\{\vec{u},\phi\} = &\frac{1}{2} \mathcal{Q}_{00}\{\vec{U}_{00}, \phi\} 
+ \sum_{n\geq 1} \mathcal{Q}_{0n}\{\vec{U}_{0n}, \phi\} \nonumber \\
&+\sum_{m\geq 1}\sum_{n\in\Z} \mathcal{Q}_{mn}\{\vec{U}_{mn}, \phi\}.
\end{align}
Since among the allowed fields $\vec{u}$ are those defined by a single pair of
 wavenumbers $(\alpha_m,\beta_n)$, the spectral
 constraint~\eqref{E:SpectralConstraint} is equivalent to requiring that each
 of the $\mathcal{Q}_{mn}$'s be positive semidefinite. The optimal upper bound
 on $\Ce$ is then achieved by minimizing the rescaled functional
\begin{equation}
\label{E:ObjectiveFunctional}
\mathcal{B}\{\phi\} = \frac{2}{\Gr} \int\limits_{-1}^{1}\left( 
\der{\phi}{\zeta}\right)^2 d\zeta - 2\phi(1)
\end{equation}
subject to the sequence of constraints $\mathcal{Q}_{mn}\geq 0$. Moreover,
 elementary functional estimates may be used to show that
\begin{equation}
\label{E:FourierEstimate}
\mathcal{Q}_{mn}%\{\vec{U}_{mn}, \phi\} 
\geq \int\limits_{-1}^{1} \left( \alpha_m^2 + \beta_n^2 
- 2 \norm{\der{\phi}{\zeta}}_\infty\right)
\norm{\vec{U}_{mn}}^2 d\zeta,
\end{equation}
where $\norm{\cdot}_\infty$ denotes the usual $L^\infty$ norm.
Hence, for a candidate background field $\phi$, only the finite set of
 wavenumbers such that 
\begin{equation}
\alpha_m^2+\beta_n^2\leq 2 %\|{\der{\phi}{\zeta}}\|_{\infty}
\norm{\der{\phi}{\zeta}}_\infty
\end{equation}
needs to be considered to show that~\eqref{E:SpectralConstraint} holds.

%%%%%%%%%%%%%%%%%%%%%%%% 
%    COMPUTATIONAL STRATEGY
%%%%%%%%%%%%%%%%%%%%%%%%
\section{Computational strategy \label{S:ComputationalStrategy}}

The principal difficulty in minimizing $\mathcal{B}\{\phi\}$ is to impose the
  non-negativity of each $\mathcal{Q}_{mn}$. The main idea behind our strategy
 is that the non-negativity of such $\phi$-dependent quadratic forms is the
 infinite dimensional equivalent of a \textit{linear matrix inequality} (LMI). 

An LMI is a condition in the form
 \begin{equation}
 \label{E:LMIdef}
 \vec{Q}(\gvec{\gamma}) := \vec{Q}_0 + \sum_{i=1}^{q} \vec{Q}_i \gvec{\gamma}_i 
\succeq 0,
 \end{equation}
 where $\gvec{\gamma}\in\R^q$ is the variable and $\vec{Q}_0, \hdots, \vec{Q}_q 
\in \R^{s\times s}$ are symmetric matrices. The notation ``$\succeq 0$" 
signifies that $\vec{Q}(\gvec{\gamma})$ is positive semidefinite, that is
$\vec{x}^T \vec{Q}(\gvec{\gamma}) \vec{x}\geq 0$ for all $\vec{x}\in\R^s$ 
(equivalently, the eigenvalues of $\vec{Q}(\gvec{\gamma})$ are non-negative).
 Note that any matrix $ \vec{Q}(\gvec{\gamma})$ whose entries are affine
 with respect to $\gvec{\gamma}$ can be written in form~\eqref{E:LMIdef}.

It can be verified that the LMI~\eqref{E:LMIdef} is a convex constraint on 
$\gvec{\gamma}$, meaning that the set $\mathcal{F}:=\{ \gvec{\gamma}\in\R^q \,\vert\, 
\vec{Q}(\gvec{\gamma}) \succeq 0 \}$ is convex: if two vectors $\gvec{\mu}, 
\gvec{\nu} \in \R^q$ satisfy~\eqref{E:LMIdef}, then so does the vector $\lambda 
\gvec{\mu}+ (1-\lambda)\gvec{\nu}$ for all $0\leq \lambda \leq 1$. 
We call $\mathcal{F}$ the \textit{feasible set} of~\eqref{E:LMIdef}.
For more details on LMIs, we refer the reader to~\cite{Boyd1994,Boyd2004}.

\begin{figure}[t]
\includegraphics[width=0.3\textwidth]{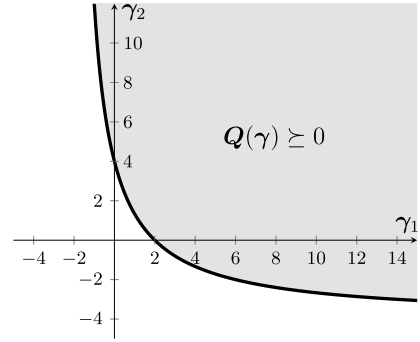}
\caption{\label{F:FeasSetExample} The shaded region indicates the
feasible set of the LMI~\eqref{E:LMIExample}, given by the points 
$\gvec{\gamma}=(\gvec{\gamma}_1,\gvec{\gamma}_2)$  satisfying both inequalities
in~\eqref{E:LMIexFeasSet}. The thick black 
line indicates the boundary of the feasible set, given by the curve 
$\gvec{\gamma}_2 = (8-4\gvec{\gamma}_1)/(2+\gvec{\gamma}_1)$.}
\end{figure}

\textbf{Example.} The condition
\begin{equation}
\label{E:LMIExample}
\vec{Q}(\gvec{\gamma}) := \begin{bmatrix}
 \gvec{\gamma}_1 +  \gvec{\gamma}_2 - 2 &  \gvec{\gamma}_2 \\
  \gvec{\gamma}_2 & 4+ \gvec{\gamma}_2
  \end{bmatrix}
  \succeq 0
 \end{equation}
 is a $2\times 2$ LMI for $ \gvec{\gamma} \in \R^2$, and can 
 be written in form~\eqref{E:LMIdef} with
 \[
 \vec{Q}_0 = \begin{bmatrix}  - 2 &  0 \\ 0 & 4  \end{bmatrix}, \quad
 \vec{Q}_1= \begin{bmatrix}  1 &  0 \\ 0 & 0  \end{bmatrix},\quad
 \vec{Q}_2 = \begin{bmatrix}  1 & 1 \\ 1 & 1  \end{bmatrix}.
 \] 
For this simple example, the feasible set $\mathcal{F}$ 
of~\eqref{E:LMIExample} can be determined analytically
by requiring that the eigenvalues of $\vec{Q}(\gvec{\gamma})$ are non-negative.
It may be verified that $\mathcal{F}$ is described by the inequalities 
 \begin{equation}
 \label{E:LMIexFeasSet}
 \begin{gathered}
 \gvec{\gamma}_1+2\gvec{\gamma}_2+2\geq 0,\\
4\gvec{\gamma}_1 + 2\gvec{\gamma}_2 +\gvec{\gamma}_1\gvec{\gamma}_2 - 8 \geq 0.
\end{gathered}
\end{equation}
For example,  the vectors $\gvec{\gamma}=(4,-1)^T$ 
 and $\gvec{\gamma}=(2,0)^T$ satisfy the LMI,  while 
$\gvec{\gamma}=(0,0)^T$ does not. As illustrated in Figure~\ref{F:FeasSetExample}, 
$\mathcal{F}$ is convex. 
\hfill $\blacksquare$

An optimization problem with a linear objective function subject to
 linear equalities and LMIs, i.e. in the form
\begin{equation}
\begin{gathered}
\min_{\gvec{\gamma}} \quad\vec{c}^T\gvec{\gamma} \\
\begin{aligned}
\text{s.t.} \quad \vec{A}\gvec{\gamma} + \vec{b} &= 0,\\
\vec{Q}(\gvec{\gamma}) &\succeq 0,
\end{aligned}
\end{gathered}
\end{equation} 
where $\vec{c}\in\R^{q}$ is the cost vector, 
$\vec{A}\in\R^{p\times q}$ and $\vec{b}\in \R^{p}$ define $p$ equality
 constraints and $\vec{Q}(\gvec{\gamma})$ is as in~\eqref{E:LMIdef}, 
 is known as a \textit{semidefinite program}  (SDP). Note that
 linear inequalities can be seen as one dimensional LMIs, and that multiple
 LMIs can always be combined into a single LMI~\cite{Boyd1994,Boyd2004}, 
 so the above form is general. 

SDPs can be solved efficiently using well-established
 algorithms~\cite{Boyd1994,Vandenberghe1996,Boyd2004}. Consequently, our 
strategy is to exploit the close relationship between the spectral 
constraint~\eqref{E:SpectralConstraint} and LMIs, and rewrite 
the variational problem for the minimization of $\mathcal{B}\{\phi\}$
 as an SDP. To this end, we will parametrize  the 
background field $\phi$ using a finite set of parameters 
(corresponding to the optimization variable $\gvec{\gamma}$ in the
 above generic SDP), and 
enforce each functional inequality $\mathcal{Q}_{mn}\geq0$ 
using sufficient conditions in the form of LMIs. This follows and extends
 the ideas already proposed by the authors in~\cite{Fantuzzi2015}. 
%
%Such LMIs will be derived by expanding all components of $\vec{U}_{mn}$ 
% with Legendre series~\cite{Zeidler1995,Jackson1930}, and using the 
% orthogonality of the Legendre polynomials with respect to the usual $L^2$
%  inner product to derive a matrix representation of the infinite
% dimensional problem. 
% The use of the Legendre series instead of the more common
% Fourier series is motivated by the need of representing functions whose BCs
% are not periodic. We also note that despite their attractive numerical properties, 
% Chebyshev polynomials do not suit our purposes because they are only 
% orthogonal  with respect to the weight $(1-\zeta)^{-1/2}$.
% Some key results on Legendre expansions are reported in
% Appendix~\ref{A:LegSeriesProperties}.

%%%%%%%%%%%%%%%%%%%%%%%% 
%    2D BOUNDS
%%%%%%%%%%%%%%%%%%%%%%%%
\section{Bounds for the two dimensional flow \label{S:2DBounds}}

The bounding problem for the two dimensional flow is obtained from
 Sections~\ref{SS:BoundingProblem}--\ref{SS:SCFourier} by neglecting the $y$ direction. We will
 also drop the suffix $n$ from all variables and functionals for simplicity. 

Equations~\eqref{E:FourierIncompressibility}--\eqref{E:FourierBCs} imply that
 $W_0=0$, so $\mathcal{Q}_0\geq 0$ for any choice of $\phi$. When $m\neq 0$,
 instead, we can use the incompressibility 
 condition~\eqref{E:FourierIncompressibility} to rewrite $U_m$ in terms of
 $W_m$ and hence express each $\mathcal{Q}_{m}$ as
\begin{align}
\mathcal{Q}_m = \int\limits_{-1}^{1}&\left[ \frac{16}{\alpha_m^2}
\abs{\der{^2W_m}{\zeta^2}}^2  + 8\abs{\der{W_m}{\zeta}}^2  
+ \alpha_m^2 \abs{W_m}^2  \right.
\nonumber\\
&\left.
- \frac{8}{\alpha_m} \der{\phi}{\zeta}\,\Im\left( 
\der{W_m}{\zeta} W_m^*\right) \right] d\zeta,
\label{E:SpectralConstraint2DFourier}
\end{align}
where the complex Fourier amplitudes $W_m(\zeta)$ satisfy the BCs
\begin{equation}
\label{E:2DBCs}
W_m(-1)= W_m(1)= \der{W_m}{\zeta}\Big\vert_{-1}= 
\der{^2W_m}{\zeta^2}\Big\vert_1 = 0.
\end{equation}

The requirement that the Fourier modes combine into a real-valued velocity
 perturbation means that $\mathcal{Q}_{-m}=\mathcal{Q}_m$, so we can restrict
 the attention to positive $m$'s. Moreover,~\eqref{E:FourierEstimate}
 guarantees that for a given choice of background field it suffices to
 consider $m$ up to the critical value
\begin{equation}
\label{E:FourierTruncation}
m_c(\phi) := \left\lfloor \frac{\Gamma_x}{\pi} \sqrt{\frac{1}{2}
\norm{\der{\phi}{\zeta}}_\infty}\right\rfloor,
\end{equation} 
where $\lfloor\cdot\rfloor$ denotes the integer part of the argument.

After rescaling the BCs for $\phi$ in~\eqref{E:BPBC}, the optimal bounds on
 $\Ce$ are determined by the solution of the variational problem
\begin{equation}
\label{E:OptProblem2D}
\begin{aligned}
&&\min_{\phi} \quad \mathcal{B}\{\phi\}& \\
\text{s.t.}  &&\mathcal{Q}_m\{W_m,\phi\} \geq0&, && 1\leq m \leq m_c(\phi), \\
&&\phi(-1)=0&, \\
&&\der{\phi}{\zeta}\Big\vert_1 = \frac{Gr}{2}&. 
\end{aligned}
\end{equation}

%%%%%%%%%%%%%%%%%%%%%%%% 
%    PARAMETRIZATION OF BACKGROUND FIELD
%%%%%%%%%%%%%%%%%%%%%%%%
\subsection{Parametrization of the background field 
\label{S:BFparametrisation}}

The first step to rewrite~\eqref{E:OptProblem2D} as an SDP is to parametrize
 the background field in terms of a finite number of decision variables. While
 the optimal $\phi$ cannot generally be described exactly with a finite
 dimensional parametrization, it can be approximated arbitrarily accurately by
 a polynomial of sufficiently high degree. Consequently, we restrict our
 attention to the family of background fields which are polynomials of degree
 at most $P+1$. 

Since our analysis of the spectral constraints will be based on Legendre series
 expansions, we parametrize the background field by expressing its first
 derivative as
\begin{equation}
\label{E:BPexpansion}
\der{\phi}{\zeta} = \sum_{p=0}^{P} \hat{\phi}_p\, \leg_p(\zeta),
\end{equation}
where $\leg_p(\zeta)$ is the Legendre polynomial of degree $p$.

The vector of Legendre coefficients
 $\gvec{\hat{\phi}}= ( \hat{\phi}_0,...,\hat{\phi}_P)^T$
 must be chosen so as to impose the correct BCs on $\phi$. The condition
 $\phi(-1)=0$ can always be enforced by an appropriate choice of integration
 constant, while since $\leg_p(1)=1$~\cite{Agarwal2009} the condition at
 $\zeta=1$ becomes
\begin{equation}
%\der{\phi}{\zeta}\Big\vert_{1} = 
\sum_{p=0}^{P} \hat{\phi}_p = \vec{1}^T\gvec{\hat{\phi}} = \frac{\Gr}{2},
\end{equation} 
where $\vec{1}\in \R^{P+1}$ is a column vector of ones.

Finally, since $\norm{\leg_p}_\infty=1$ for all $p$ we have
 the useful estimate
\begin{equation}
\label{E:InfNormBPEstimate}
\norm{\der{\phi}{\zeta}}_{\infty} = \max_{\zeta\in[-1,1]} 
\abs{ \sum_{p=0}^{P} \hat{\phi}_p \leg_p(\zeta)} \leq 
%\sum_{p=0}^{P} \abs{\hat{\phi}_p} =
\|\gvec{\hat{\phi}}\|_1,
\end{equation}
where $\norm{\cdot}_1$ denotes the usual $l^1$ norm.

\subsection{Formulation of a linear cost function 
\label{SS:2DLinearCostFunction}}

In order to formulate~\eqref{E:OptProblem2D} as a standard SDP,
 we need to replace the quadratic objective functional $\mathcal{B}$
 by an equivalent linear cost function. 
The orthogonality of the Legendre polynomials allows us to rewrite 
\begin{equation}
\int\limits_{-1}^{1}\left( \der{\phi}{\zeta}\right)^2 d\zeta = 
\gvec{\hat{\phi}}^T \vec{B} \gvec{\hat{\phi}}, 
\end{equation}
where $\vec{B}\in\R^{(P+1)\times(P+1)}$ is defined as
\begin{equation}
\vec{B}_{rs} = \frac{2\delta_{rs}}{2r+1}, \qquad 0\leq r,s \leq P
\end{equation}
and $\delta_{rs}$ is the usual Kroenecher delta.
Moreover, using the results of Appendix~\ref{A:LegSeriesProperties} we have
 $\phi(1) = \phi(-1) + 2\hat{\phi}_0$. Applying the boundary condition 
$\phi(-1)=0$, the objective functional $\mathcal{B}$ becomes
\begin{equation}
\mathcal{B}\{\phi\} = \frac{2}{Gr} \gvec{\hat{\phi}}^T \vec{B} 
\gvec{\hat{\phi}} - 4\hat{\phi}_0.
\end{equation}

Using a standard trick of convex optimization, we introduce an additional
 decision variable $\eta$ (called a \textit{slack variable}) and minimize
\begin{equation}
\label{E:LinObjFunction}
\mathcal{B}_{\text{lin}}(\gvec{\hat{\phi}},\eta) := 
\frac{2}{Gr} \eta - 4\hat{\phi}_0
\end{equation}
instead of $\mathcal{B}$, subject to the additional constraint that 
$\eta \geq \gvec{\hat{\phi}}^T \vec{B} \gvec{\hat{\phi}}$.
Since $\vec{B}$ is invertible and its inverse is positive definite,
 Schur's complement condition~\cite{Boyd2004} guarantees that
\begin{equation}
\label{E:SchurComplement}
\vec{S}(\gvec{\hat{\phi}},\eta) := \begin{bmatrix}
\vec{B}^{-1} & \gvec{\hat{\phi}} \\ \gvec{\hat{\phi}}^T & \eta
\end{bmatrix} \succeq 0 
\quad \Longleftrightarrow \quad
\eta \geq \gvec{\hat{\phi}}^T \vec{B} \gvec{\hat{\phi}}.
\end{equation}
Hence, the additional inequality constraint can be expressed as the LMI
 $\vec{S}(\gvec{\hat{\phi}},\eta)\succeq 0$.

\subsection{Rigorous finite dimensional relaxation of $\mathcal{Q}_m$ using
 Legendre expansions \label{SS:LMIRelaxation2D}}

We now turn to the core problem of deriving a rigorous matrix representation
 for each of the constraints $\mathcal{Q}_m\geq 0$, which will allow us 
 to enforce the spectral constraint using LMIs.
 As in~\cite{Fantuzzi2015},
 the analysis is based on the application of orthogonal series expansions. The need to 
 enforce BCs other than periodicity prevents us from using the conventional
 Fourier series. As our analysis relies on $L^2$ orthogonality of the basis
  functions, we will use Legendre series 
  expansions~\cite{Jackson1930,Zeidler1995} (note that despite their
   attractive numerical properties, the more commonly used
 Chebyshev polynomials do not suit our purposes because they are only 
 orthogonal  with respect to the weight $\sqrt{1-\zeta^2}$).
Some key results on Legendre expansions are reported in
Appendix~\ref{A:LegSeriesProperties}.
The details of the following analysis are quite technical, and will be omitted
 to keep the focus on our main objective --- formulating an SDP
  whose solution gives a feasible
 background field for~\eqref{E:OptProblem2D}. LMI relaxations of integral
 inequalities using Legendre series will be thoroughly discussed in a future
 publication~\cite{Fantuzzi2015a}.
Moreover, for notational neatness, we will drop the suffix $m$ from
 $\mathcal{Q}_m$, $W_m$ and $\alpha_m$; it should be understood that
 the following analysis holds for each individual $m\geq1$.

Since the function $W$ represents the amplitude of a Fourier mode of
 a velocity perturbation, we assume it has enough regularity such that $W$,
 $\der{W}{\zeta}$ and $\der{^2W}{\zeta^2}$ can be expanded with the uniformly
 convergent Legendre series~\cite{Jackson1930,Zeidler1995}
\begin{subequations}
\label{E:LegExp}
\begin{align}
W  &= \sum_{n=0}^{\infty} \hat{w}_n \leg_n(\zeta),\\
\der{W}{\zeta}&=\sum_{n=0}^{\infty} \hat{w}'_n \leg_n(\zeta), \\
\der{^2W}{\zeta^2}&=\sum_{n=0}^{\infty} \hat{w}''_n \leg_n(\zeta).
\end{align}
\end{subequations}
Since $W$ is complex valued, so are the Legendre coefficients $\hat{w}_n$,
 $\hat{w}_n'$ and $\hat{w}_n''$. In addition, the results of
 Appendix~\ref{A:LegSeriesProperties} and the BCs at $\zeta=-1$
 from~\eqref{E:2DBCs} imply that the Legendre coefficients are related
 by the compatibility conditions 
\begin{subequations}
\label{E:IntegrationRule}
\begin{equation}
\label{E:IntegrationRule_a}
\begin{aligned}
\hat{w}_0 &= \hat{w}'_0 -\frac{\hat{w}'_1}{3}, \\
\hat{w}_n &=  \frac{\hat{w}'_{n-1}}{2n-1} -\frac{\hat{w}'_{n+1}}{2n+3},
\qquad n\geq 1,
\end{aligned}
\end{equation}
and
\begin{equation}
\label{E:IntegrationRule_b}
\begin{aligned}
\hat{w}'_0 &= \hat{w}''_0 -\frac{\hat{w}''_1}{3}, \\
 \hat{w}'_n &=  \frac{\hat{w}''_{n-1}}{2n-1} -\frac{\hat{w}''_{n+1}}{2n+3},
\qquad n\geq 1.
\end{aligned}
\end{equation}
\end{subequations}

Rather than truncating the Legendre series of $W$ after $N$ terms to obtain
 an approximation of $\mathcal{Q}$, we consider the full infinite dimensional
 quadratic form by defining the remainder functions
\begin{subequations}
\label{E:LegtruncatedExp}
\begin{align}
\label{E:LegtruncatedExp_a} \tilde{w}_0(\zeta) &:=  
\sum_{n=N+1}^{\infty} \hat{w}_n \leg_n(\zeta), \\
\label{E:LegtruncatedExp_b} \tilde{w}_1(\zeta) &:=  
\sum_{n=N+2}^{\infty} \hat{w}'_n \leg_n(\zeta), \\
\label{E:LegtruncatedExp_c} \tilde{w}_2(\zeta) &:=  
\sum_{n=N+P+4}^{\infty} \hat{w}''_n \leg_n(\zeta).
\end{align}
\end{subequations}
The lower limit in~\eqref{E:LegtruncatedExp_b} is motivated
 by~\eqref{E:IntegrationRule_a}, while the lower limit $N+P+4$
 in~\eqref{E:LegtruncatedExp_c} is chosen for convenience; this will become
 apparent in Appendix~\ref{A:LMIrelaxation}.

\begin{figure}[t]
\includegraphics[width=0.45\textwidth, trim=0cm 0cm 0cm 0.5cm]{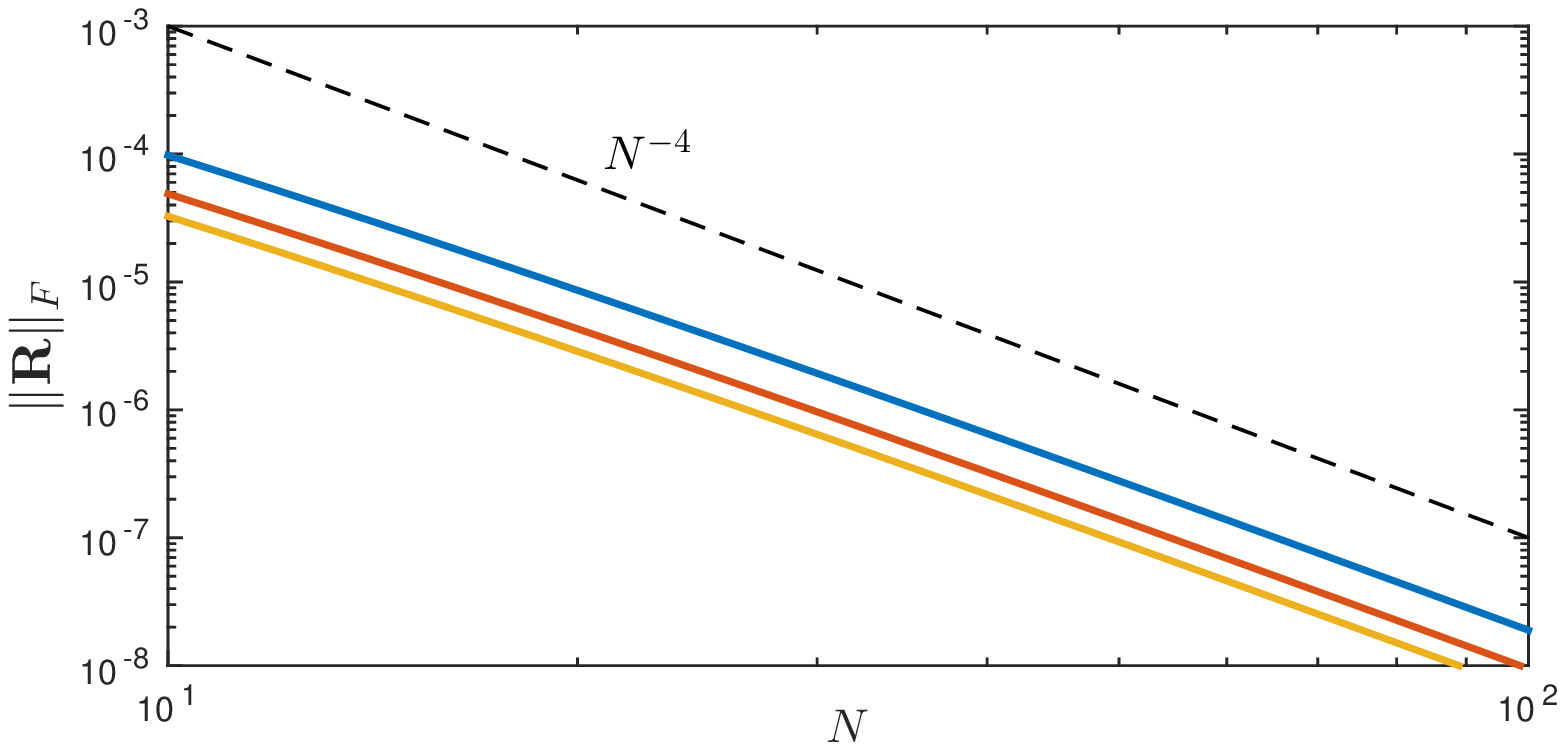}
\includegraphics[width=0.45\textwidth, trim=0cm 0.5cm 0cm 0cm]{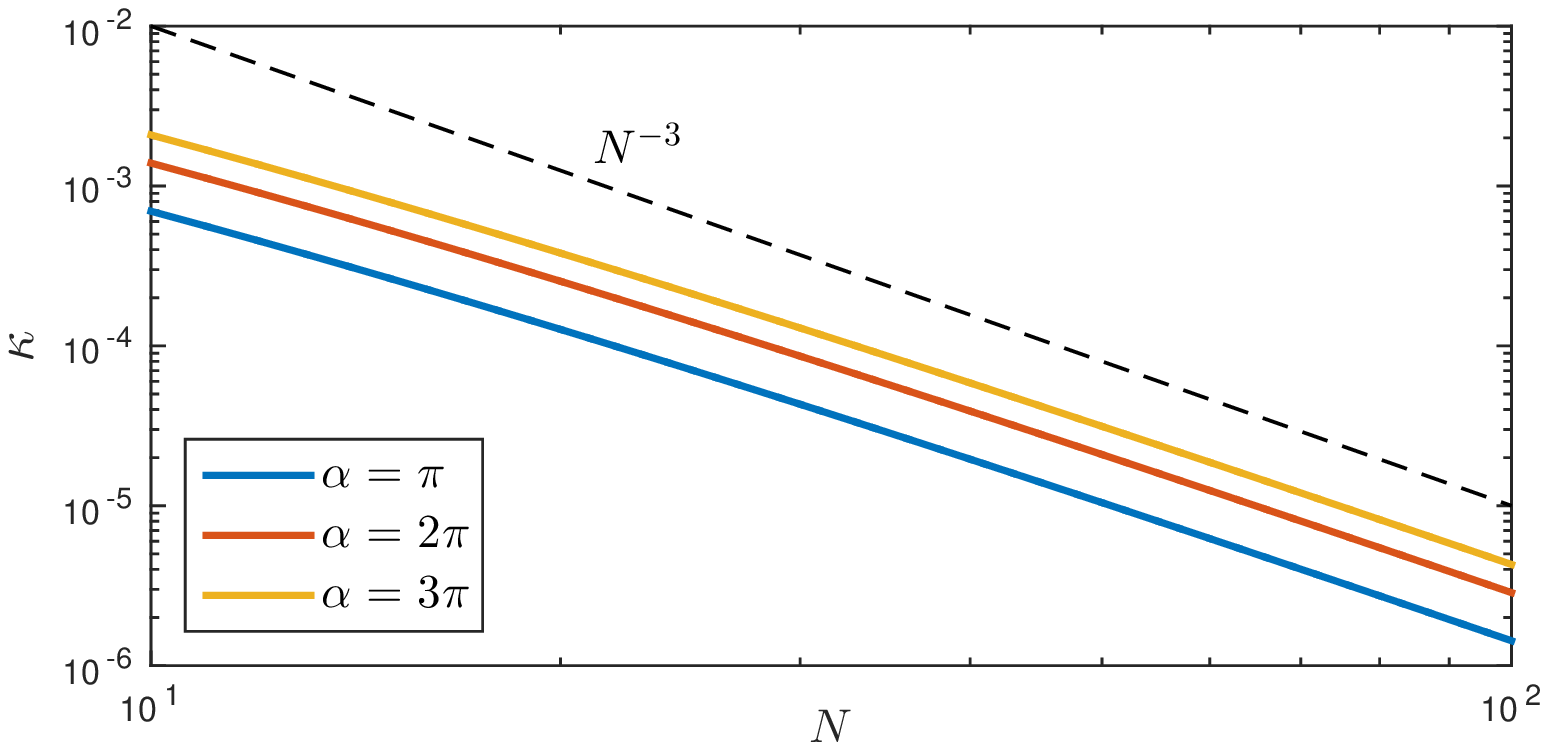}
\caption{\label{F:NormRandKappa} (Color Online) Decay of the Frobenius
 norm of the matrix $\vec{R}$ (top) and the constant $\kappa$ (bottom)
 for selected wavenumbers $\alpha$.}
\end{figure}

In Appendix~\ref{A:LMIrelaxation}, we show that there exist a real, symmetric,
 positive definite matrix $\vec{Q}_1 \in\R^{(N+P+4)\times(N+P+4)}$, and a real
(but \emph{not} symmetric) matrix 
$\vec{Q}_2(\gvec{\hat{\phi}})\in\R^{(N+P+4)\times(N+P+4)}$, 
whose entries are linear in $\gvec{\hat{\phi}}$, such that
\begin{align}
\mathcal{Q}\{W,\phi\} = &\left(\vec{\hat{w}}''\right)^\dagger
 \vec{Q}_1 \vec{\hat{w}}'' -  \Im \left[ \left(\vec{\hat{w}}''\right)^\dagger
 \vec{Q}_2(\gvec{\hat{\phi}}) \vec{\hat{w}}''\right] 
\nonumber\\
&+\mathcal{R}\{\tilde{w}_0,\tilde{w}_1,\tilde{w}_2,\phi\}.
\label{E:QExpansion1}
\end{align}
Here, $(\cdot)^\dagger$ denotes conjugate transposition, 
$\vec{\hat{w}}''\in\C^{N+P+4}$ is the column vector containing the Legendre
 coefficients $\hat{w}''_0$, ..., $\hat{w}''_{N+P+3}$ and
\begin{align}
\label{E:Rdef}
\mathcal{R}\{\tilde{w}_0,\tilde{w}_1,\tilde{w}_2,\phi\} := 
\int\limits_{-1}^{1}&\left[ \frac{16}{\alpha^2}\abs{\tilde{w}_2}^2
 + 8\abs{\tilde{w}_1}^2  + \alpha^2 \abs{\tilde{w}_0}^2 \right.
\nonumber\\
&\left.- \frac{8}{\alpha} \der{\phi}{\zeta}\,\Im\left( 
\tilde{w}_1 \tilde{w}_0^*\right) \right] d\zeta.
\end{align}
We also show in Appendix~\ref{A:LowerBoundR} that
\begin{align}
\label{E:LowerBoundR}
\mathcal{R}\{\tilde{w}_0,\tilde{w}_1,\tilde{w}_2,\phi\} \geq
  &-\|\gvec{\hat{\phi}}\|_1 \left(\vec{\hat{w}}''\right)^\dagger 
\vec{R} \vec{\hat{w}}'' 
\nonumber \\
&+\frac{16}{\alpha^2} \left( 1-\kappa \|\gvec{\hat{\phi}}\|_1\right)
 \norm{\tilde{w}_2}^2_2,
\end{align}
where $\norm{\cdot}_2$ is the usual $L^2$ norm, 
$\vec{R}\in\R^{(N+P+4)\times(N+P+4)}$ is a positive definite real matrix
 whose Frobenius norm $\norm{\vec{R}}_F$ scales as  $\mathcal{O}(N^{-4})$,
 and $\kappa$ is a positive constant scaling as $\mathcal{O}(N^{-3})$ (see
 Figure~\ref{F:NormRandKappa}). A more in-depth discussion of the properties of
 $\vec{Q}_1$, $\vec{Q}_2$, $\vec{R}$ and $\kappa$ will be given
 elsewhere~\cite{Fantuzzi2015a}.
Considering the real and imaginary parts of $\vec{\hat{w}}''$ explicitly
and defining
\begin{equation}
\label{E:MphiDefinition}
\vec{M}(\gvec{\hat{\phi}}) := \mathrm{sym}\left( \begin{bmatrix}
\displaystyle\vec{Q}_1 - \|\gvec{\hat{\phi}}\|_1\vec{R} &
 -\vec{Q}_2(\gvec{\hat{\phi}}) \\
 \vec{Q}_2(\gvec{\hat{\phi}}) & \displaystyle\vec{Q}_1 
- \|\gvec{\hat{\phi}}\|_1\vec{R}
\end{bmatrix} \right),
\end{equation}
where $\mathrm{sym}(\cdot)$ denotes the symmetric part of a matrix, 
we obtain the rigorous lower bound
\begin{equation}
\mathcal{Q}\{W,\phi\} \geq \Psi\{\vec{\hat{w}}'',\tilde{w}_2,\phi\},
\label{E:SpConstrLB1}
\end{equation}
where
\begin{align}
\label{E:PsiDefinition}
\Psi\{\vec{\hat{w}}'',\tilde{w}_2,\phi\} := &\begin{bmatrix}
\Re\left(\vec{\hat{w}}''\right) \\  \Im\left(\vec{\hat{w}}''\right)
\end{bmatrix}^T
\vec{M}(\gvec{\hat{\phi}})
\begin{bmatrix}
\Re\left(\vec{\hat{w}}''\right) \\  \Im\left(\vec{\hat{w}}''\right)
\end{bmatrix}
\nonumber\\
&+ \frac{16}{\alpha^2}\left( 1-\kappa \,\|\gvec{\hat{\phi}}\|_1\right)
 \norm{\tilde{w}_2}^2_2.
 \end{align}
% 
%We remark that the off-diagonal blocks do \emph{not} cancel when taking the 
%symmetric part in~\eqref{E:MphiDefinition} because $\vec{Q}_2$ is not symmetric.
%
Note that this lower bound holds for all functions $W$ that satisfy the BCs at 
$\zeta=-1$ in~\eqref{E:2DBCs}. Consequently, the positivity of $\mathcal{Q}$ is 
proven for any $W$ satisfying the boundary conditions at $\zeta=-1$ if we 
enforce
\begin{subequations}
\label{E:StrongSuffCond}
\begin{gather}
\vec{M}(\gvec{\hat{\phi}})\succeq 0, \\
1 - \kappa \| \gvec{\hat{\phi}} \|_1 \geq 0.
\end{gather}
\end{subequations}

However, $\mathcal{Q}$ needs only be positive for any $W$ satisfying 
\textit{all} BCs in~\eqref{E:2DBCs}, which is a weaker statement. 
Recalling that $\leg_p(\pm1)=(\pm1)^p$ for all $p$, with the help
 of~\eqref{E:IntegrationRule} and letting $\vec{1}\in\R^{N+P+4}$ 
 be a column vector of ones, the BCs at
 $\zeta = 1$ from~\eqref{E:2DBCs} can be expressed as
\begin{subequations}
\label{E:AdmissibilityBCs}
\begin{gather}
\label{E:AdmissibilityBCs_1}
\hat{w}''_0 -\frac{\hat{w}''_1}{3} = 0, \\
\label{E:AdmissibilityBCs_2} 
%\sum_{n=0}^{N+P+3} \hat{w}''_n + \tilde{w}_2(1)= 0.
\vec{1}^T \vec{\hat{w}}''+ \tilde{w}_2(1)= 0.
\end{gather}
\end{subequations}

The first of these conditions can be imposed by letting
\begin{equation}
\label{E:BC1representation}
\vec{\hat{w}}'' = \vec{A} \gvec{\hat{\omega}},
\end{equation}
where
$$
\gvec{\hat{\omega}} := \begin{pmatrix}
\hat{w}''_1 \\ \vdots \\ \hat{w}''_{N+P+3}
\end{pmatrix}, \qquad 
\vec{A} := \begin{bmatrix}
^1/_3 & 0  & \hdots & 0 \\
    1 & 0  & \hdots & 0 \\
    0 & 1  & \hdots & 0 \\
\vdots&  & \ddots &    \vdots\\
 0 &  \hdots &   0& 1
\end{bmatrix}.
$$
Substituting this representation in~\eqref{E:PsiDefinition} and defining
\begin{equation}
\vec{Q}(\gvec{\hat{\phi}}) :=
\begin{bmatrix}
\vec{A} & \vec{0} \\ \vec{0} & \vec{A}
\end{bmatrix}^T
\vec{M}(\gvec{\hat{\phi}}) 
\begin{bmatrix}
\vec{A} & \vec{0} \\ \vec{0} & \vec{A}
\end{bmatrix}
\end{equation}
we obtain
\begin{align}
\label{E:PsiWithBC}
\Psi\{\vec{\hat{w}}'',\tilde{w}_2,\phi\} =
&\begin{bmatrix}
\Re\left(\gvec{\hat{\omega}}\right) \\  \Im\left(\gvec{\hat{\omega}}\right)
\end{bmatrix}^T
\vec{Q}(\gvec{\hat{\phi}})
\begin{bmatrix}
\Re\left(\gvec{\hat{\omega}}\right) \\  \Im\left(\gvec{\hat{\omega}}\right)
\end{bmatrix}
\nonumber \\
&+\frac{16}{\alpha^2}\left( 1-\kappa \,\|\gvec{\hat{\phi}}\|_1\right)
 \norm{\tilde{w}_2}^2_2.
\end{align}

Thus, the spectral constraint can be enforced via the 
sufficient finite dimensional conditions
\begin{subequations}
\label{E:2DSufficientConditions}
\begin{gather}
\label{E:SuffCondLMI}
\vec{Q}(\gvec{\hat{\phi}}) \succeq 0, \\
\label{E:SuffCondLinIneq}
1-\kappa \,\|\gvec{\hat{\phi}}\|_1 \geq 0,
\end{gather}
\end{subequations}
which are weaker than~\eqref{E:StrongSuffCond} and, consequently, allow us to 
compute a better bound.

These sufficient conditions could be weakened further 
if~\eqref{E:AdmissibilityBCs_2} could be incorporated. 
However, this is difficult to achieve since $\tilde{w}_2(1)$ does not appear 
explicitly in~\eqref{E:PsiWithBC}. 
Moreover, we show in Appendix~\ref{A:CommentsSuffCond} 
that~\eqref{E:2DSufficientConditions} must still be satisfied if $\Psi\geq 0$ 
when~\eqref{E:AdmissibilityBCs_2} holds. 
Consequently, we choose to enforce the spectral constraint via~\eqref{E:2DSufficientConditions}.

\begin{remark}\label{R:Remark1}
Since the matrix $\vec{R}$ of 
equation~\eqref{E:LowerBoundR} gives a negative definite contribution to 
$\vec{Q}$ and $\kappa$ is positive, it is not
 obvious that the conditions in~\eqref{E:2DSufficientConditions} are feasible.
 However, $\norm{\vec{R}}_F$ and $\kappa$ decay to zero as at least as fast
 as $N^{-3}$. Under the reasonable assumption that a \textit{strictly feasible}
 polynomial background field of degree $P+1$ exists
 for~\eqref{E:SpectralConstraint}, we expect that a vector 
 $\gvec{\hat{\phi}}$ satisfying~\eqref{E:2DSufficientConditions} exists when
 $N$ is sufficiently large. Moreover, although the size of $\vec{Q}$ increases
 linearly with $N$, the fast decay of $\norm{\vec{R}}_F$ and $\kappa$ means
 that in practice the required $N$ is small enough that~\eqref{E:SuffCondLMI} 
 is a numerically tractable constraint.
 \end{remark}

\begin{remark}\label{R:Remark2}
Conditions~\eqref{E:2DSufficientConditions} need to be derived and imposed for a 
range of wavenumbers $\alpha_m$, $m = 1,...,m_c$. Since $\kappa$ increases
linearly with $\alpha_m$ (equivalently, with $m$, cf. 
Appendix~\ref{A:LowerBoundR}), we expect that the truncation $N$ required to 
make~\eqref{E:SuffCondLinIneq} feasible increases with $m$. Similarly, although 
$\norm{\vec{R}}_F$ decreases linearly with $m$ (cf. 
Appendix~\ref{A:LowerBoundR}), the dominant positive definite contribution to 
$\vec{Q}(\gvec{\hat{\phi}})$ --- which comes from the the second derivative term 
in the functional $\mathcal{Q}$ --- decays as $m^{-2}$. Hence,  we again expect 
that a higher truncation $N$ will be required to make~\eqref{E:SuffCondLMI} 
feasible for large values of $m$.
 \end{remark}

\subsection{An SDP for the optimal bounds \label{SS:SDP2D}}

The conditions in~\eqref{E:2DSufficientConditions} are not LMIs due to the
 appearance of absolute values in $\|\gvec{\hat{\phi}}\|_1$, but can be recast
 as LMIs by introducing a vector $\vec{t}= (t_0,...,t_{P})^T$ of slack
 variables, replacing $\|\gvec{\hat{\phi}}\|_1$ with $\vec{1}^T\vec{t}=\sum_{p=0}^{P}t_p$ and
 introducing $2P+2$ inequality constraints of the form
\begin{equation}
\hat{\phi}_j -t_j\leq 0, \qquad \hat{\phi}_j +t_j\geq 0.
\end{equation}
As a result, each of the functional inequalities $\mathcal{Q}_m\geq 0$ 
in~\eqref{E:OptProblem2D} can be replaced by the LMI 
$\vec{Q}_m(\gvec{\hat{\phi}},\vec{t})\succeq 0$ and the linear inequality
 $1-\kappa_m \vec{1}^T\vec{t}\geq 0$ (where $\vec{Q}_m$ and $\kappa_m$ are
 derived as in Section~\ref{SS:LMIRelaxation2D} for each $m$). Together with 
 equations~\eqref{E:LinObjFunction}--\eqref{E:SchurComplement}, this means
 that bounds on $\Ce$ can be computed at each $\Gr$ after
 solving the SDP
\begin{equation}
\label{E:SDP2D}
\begin{aligned}
&\min_{\eta,\gvec{\hat{\phi}},\vec{t}} \qquad 
\frac{2}{Gr} \eta - 4\hat{\phi}_0
\\
&\begin{aligned}
\text{s.t.} \qquad\qquad \vec{S}(\gvec{\hat{\phi}},\eta) \succeq 0,&\\
\vec{Q}_m(\gvec{\hat{\phi}},\vec{t})\succeq 0,&  &&&&1\leq m \leq m_c,\\
1-\kappa_m\vec{1}^T\vec{t} \geq 0,&  &&&&1\leq m \leq m_c, \\
\hat{\phi}_j - t_j \leq 0,&  &&&&0\leq j \leq P,\\
\hat{\phi}_j + t_j \geq 0,&  &&&&0\leq j \leq P, \\
\vec{1}^T\gvec{\hat{\phi}} = \frac{\Gr}{2}.
\end{aligned}
\end{aligned}
\end{equation}

Strictly speaking, the bounds computed with the solution of this SDP are not
 optimal, but only \emph{near-optimal} because they are obtained using a
 restricted class of background fields and imposing a stronger condition than
 the original spectral constraint. 
In practice, however, we can increase the parameters $N$ and $P$ until the
 solution of the SDP has converged to that of~\eqref{E:OptProblem2D} --- 
 at the expense of increasing the computational cost of the optimization. 

In addition, the analysis of Section~\ref{SS:LMIRelaxation2D} guarantees 
 that the background field constructed with the optimal $\gvec{\hat{\phi}}$
 is a feasible choice for~\eqref{E:OptProblem2D}, so the bounds computed
  at each $\Gr$ would be rigorous if the numerical roundoff errors in the solution 
of~\eqref{E:SDP2D} were tracked and carefully taken into account. 
The implementation of algorithms  to solve~\eqref{E:SDP2D} rigorously
 is beyond the scope of this work and the bounds 
presented in the following sections cannot be considered analytical results. 
However, a fully computer-assisted proof of near-optimal bounds does not seem 
beyond the reach of future work.

Finally, we remark that $m_c$ depends on $\phi$ according 
to~\eqref{E:FourierTruncation}, so the number of inequalities in the SDP is not 
known \emph{a priori}. This means that an iterative 
procedure is needed: solve the optimization using a suitable initial guess $m_0$ 
for $m_c$, calculate the correct $m_c$ with~\eqref{E:FourierTruncation} after 
the optimization, and check the full set of LMIs \emph{a posteriori}; if any 
constraints are violated, the optimization is repeated with the updated $m_c$.
We also remark that the role of $m_c$ is that of an upper bound on the largest 
critical Fourier mode --- that is the largest value of $m$  for which the 
constraints in~\eqref{E:SDP2D} are active. Of course, if one knew the exact 
critical modes \textit{a priori}, one could solve the SDP by considering only 
such modes. However, the fact that the number of inequalities in~\eqref{E:SDP2D} 
is unknown is not due to the lack of knowledge of the exact critical modes, but 
only to the dependence of $m_c$ on $\phi$. In fact, if one could find an 
explicit value for $m_c$, say in terms of the Grashoff number, the number of 
LMIs in the SDP would be well defined and no iterative procedure would be 
required.

\subsection{Numerical implementation and results \label{SS:Results2D}}

\begin{table}[b]
  \caption{Problem data, memory requirements and CPU time for 
selected SDP instances. CPU time and memory include pre- and post-processing 
routines as well as the solution of the SDP.}
 \centering
 \begin{tabular}{ccccccccc}
 \hline\hline
 $\Gamma_x$ &  $\Gr$ & $m_0$ & $P$ & $N$ & RAM (Mb) &  Time (s) & $m_c$ & 
$N_{\mathrm{checks}}$\\
 \hline
 2 & $10^3$ &5 & 19 & 100 & 7.4 & 50 &  12 & 100\\
 2 & $10^4$ &8 & 29 & 125 & 25.3 & 160 & 38 & 125\\
 2 & $10^5$ & 15 & 34 & 150 & 71.7 & 1280 & 116 & 350\\
 \hline 
 3 & $10^3$ &  5  & 19 & 100 &  7.3 & 59 & 18  & 100 \\
3 & $10^4$  & 10 & 29 & 125 &  31.6 &  188 &  57 & 125\\
3 &  $10^5$ & 25 & 34 & 150 &  119.4 &  2608 &  174 & 350 \\
\hline\hline
 \end{tabular}
  \label{T:Data2D}
 \end{table}

\begin{figure*}
\includegraphics[width=0.32\textwidth, trim=1cm 0.25cm 0cm 0.5cm]{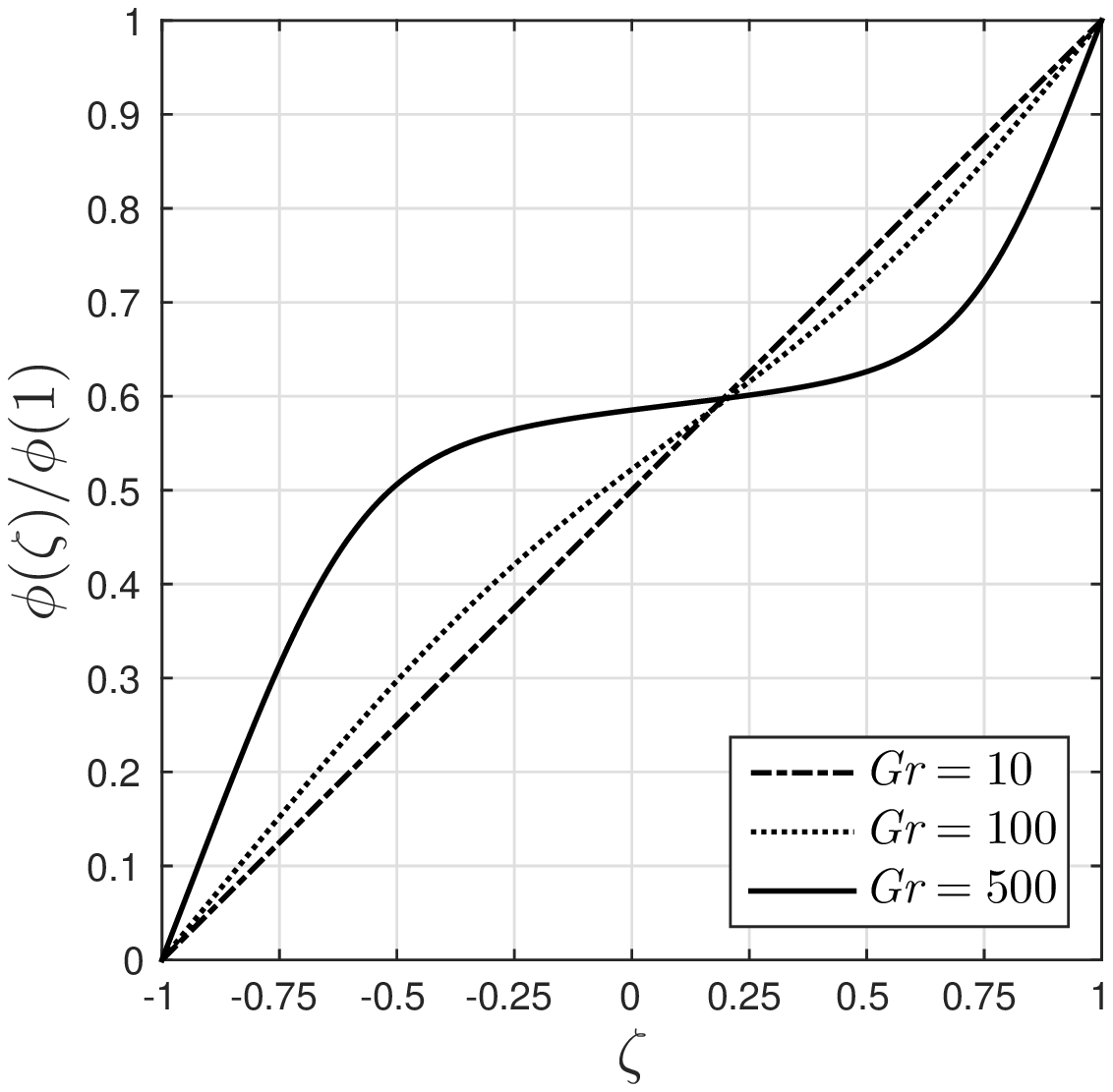}
\includegraphics[width=0.32\textwidth, trim=0cm 0.25cm 1cm 0.5cm]{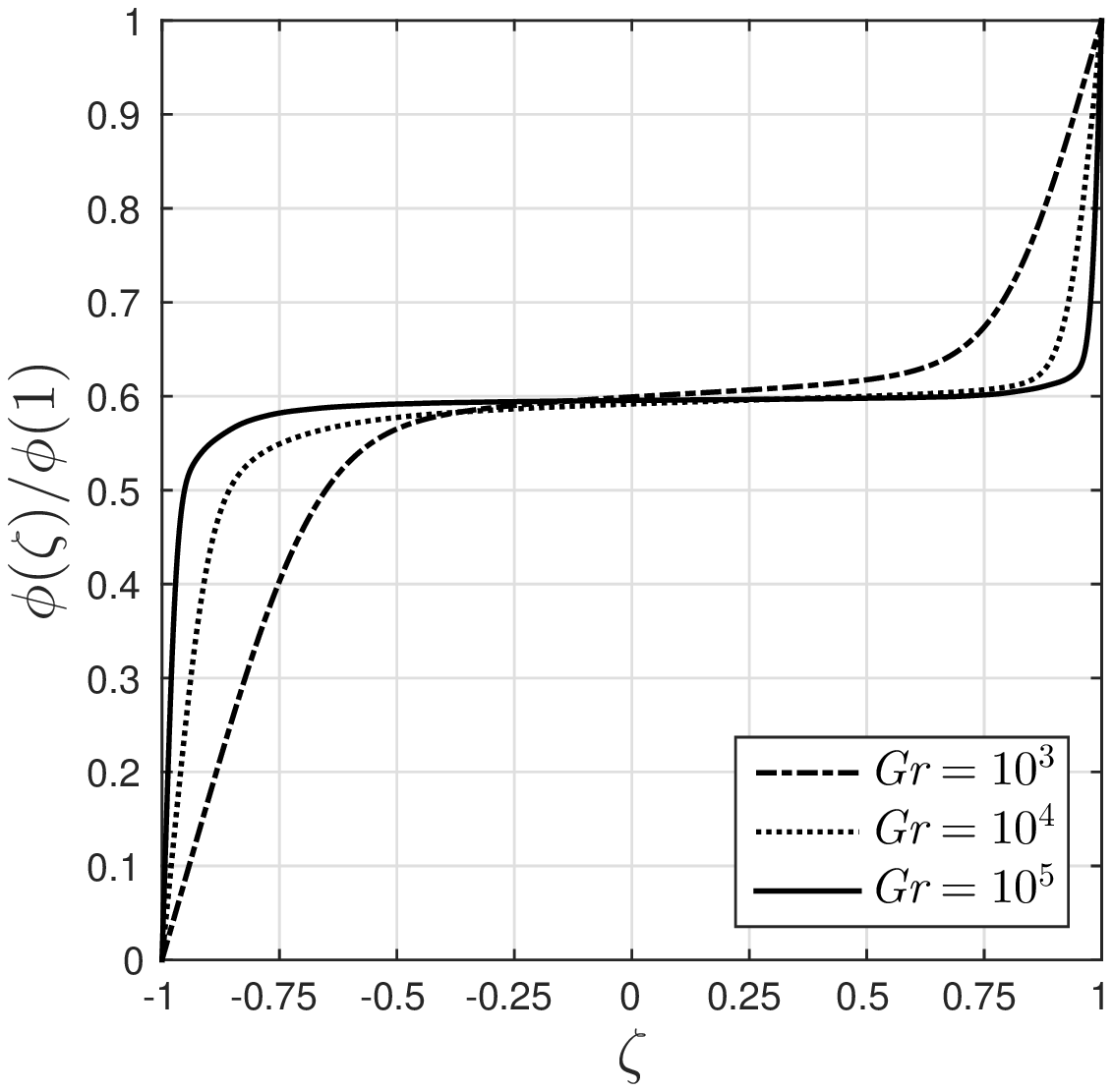}
\caption{\label{F:BP2DGammax2}Background fields normalized by their value 
at $\zeta=1$ for selected Grashoff numbers and $\Gamma_x=2$.}
\end{figure*}
\begin{figure*}
\includegraphics[width=0.75\textwidth, trim = 0cm 0.25cm 0cm 0.5cm]{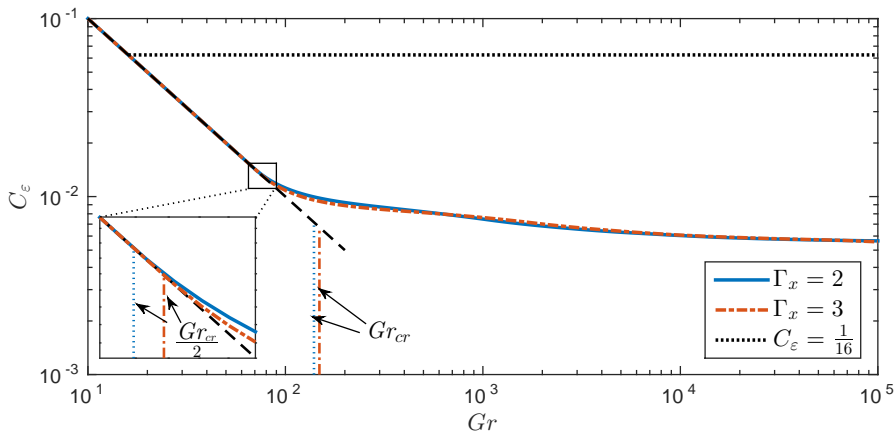}
\caption{\label{F:2DOptimalBounds} (Color Online) Optimal upper bounds 
 on $\Ce$ for $\Gamma_x=2$ and $\Gamma_x=3$, compared to the analytical bound
 $\Ce\leq\frac{1}{16}$ from~\cite{Hagstrom2014}. The laminar dissipation
 coefficient is shown as a dashed black line. Detail: as expected, the bounds
 depart from the laminar $\Ce$ at $\Gr = 0.5\Gr_{\text{cr}}$, where
 $\Gr_{\text{cr}}=139.54$ for $\Gamma_x=2$ and $Gr_{cr}=148.66$
 for $\Gamma_x=3$ (also shown).}
\end{figure*}

The SDP~\eqref{E:SDP2D} was solved in {MATLAB} using the optimization toolbox
 YALMIP~\cite{Lofberg2004} and the SDP solver SeDuMi~\cite{Sturm1999} for
 $10\leq\Gr\leq 10^{5}$ and for two values of domain aspect ratio, 
 $\Gamma_x=2$ and $\Gamma_x=3$. All computations were carried out on a desktop 
computer with an Intel Core i7 3.40GHz CPU and 16Gb of RAM.
 At each $\Gr$, we chose an initial guess $m_0$, fixed the degree $P$ of the 
 background field  and the number $N$ of Legendre modes of the 
 perturbations, and solved the SDP.  We increased $P$ and $N$ until the optimal 
bounds on $\Ce$ changed by less than approximately $1$\%. Finally, given the 
optimal background field, we computed $m_c(\phi)$ 
with~\eqref{E:FourierTruncation} and verified that $\phi$ was feasible for all 
$m\leq m_c$. 
Since in practice $m_c \gg m_0$ at large $\Gr$, we expect that these checks 
might fail for the largest values of $m$ (cf. Remark~\ref{R:Remark2}). Before 
repeating the optimization with the updated set of LMIs, we therefore try to 
validate the background field with  an increased value of $N$. In the following, 
we will refer to this value as $N_{\mathrm{checks}}$.

Details of the problem specifications, memory requirements, computation time and
\textit{a posteriori} checks for selected SDP instances are reported in 
Table~\ref{T:Data2D}. 
The reason for the disparity in the number of Legendre coefficients used for 
the background field and the perturbation is that $N$ needs to be large 
to make the negative definite 
terms in the constraints small 
enough to obtain a feasible and accurate SDP formulation 
of~\eqref{E:OptProblem2D} (cf. Remark~\ref{R:Remark1}).
Instead, the optimal $\phi$ is well resolved with modest $P$. 

Figure~\ref{F:BP2DGammax2} shows some of the optimal background fields
 obtained for $\Gamma_x=2$, normalized by their boundary value $\phi(1)$.
 Analogous results were obtained for $\Gamma_x=3$ and are not shown for
 brevity. As $\Gr$ is raised, the background fields evolve from the linear
 Couette profile, developing two boundary layers in a similar way to that
 observed by Nicodemus \textit{et al}.~\cite{Nicodemus1998} for the classical
 Couette flow. The asymmetric depth of the boundary layers reflects the 
 asymmetry of the BCs.

The bounds on $\Ce$ corresponding to the optimal background fields are
 illustrated in Figure~\ref{F:2DOptimalBounds},  along with the laminar
 dissipation coefficient and the analytical bound $\Ce\leq 1/16 = 0.0625$
 proven in~\cite{Hagstrom2014}. Although energy stability analysis 
 indicates that the laminar Couette flow is
 stable up to the critical Grashoff number $\Gr_{\text{cr}}=139.54$ for
 $\Gamma_x=2$ and $\Gr_{\text{cr}}=148.66$ for $\Gamma_x=3$ 
 (cf. Section~\ref{S:EnergyStability}), our bounds 
 deviate from the laminar value $\Ce=\Gr^{-1}$ when
 $\Gr\geq 0.5\Gr_{\text{cr}}$. This was expected, because the 
 spectral constraint of the bounding
 problem differs from that of the energy stability problem  by a factor of 2
 when letting $\phi=\vec{u}_{L}$ (cf. equations~\eqref{E:EnStabConstr}
 and~\eqref{E:SpectralConstraint} in Section~\ref{S:ShearFlowIntro}). While
 this limitation could be overcome by the addition of a balance parameter in
 the spectral constraint (see 
 e.g.~\cite{Nicodemus1998a,Wittenberg2010a,Wen2015a}),
 it allows us to check that the solution of our SDP has converged to that of the
 original bounding problem. 
 
  As predicted by the
 analytical bounds of~\cite{Hagstrom2014}, the results suggest that $\Ce$
 approaches a constant when $\Gr\to\infty$ independently of $\Gamma_x$,
 meaning that the dissipation coefficient becomes independent of the flow
 viscosity and aspect ratio. The quantitative improvement, however, is
 evident: our near-optimal bound is more than 10 times smaller than the
 analytical result at $\Gr=10^5$. Unfortunately, the limited range of $\Gr$  
does not allow us to confidently estimate an asymptotic value for $\Ce$ (see 
Section~\ref{SS:ComputIssues} for further comments on the 
computational issues at large $\Gr$). 
Moreover, we could not compare our bounds to values of $\Ce$ extracted from 
experiments or direct numerical simulations because, to the best of our 
knowledge, such data are not available. Although it is unlikely that there 
exists a solution to the Navier--Stokes equation whose associated dissipation 
coefficient equals our bounds~\cite{Tang2004}, performing this comparison 
remains in the interest of future research.

\begin{figure}[t]
\includegraphics[width=0.47\textwidth, trim=1cm 0cm 1cm 0cm]{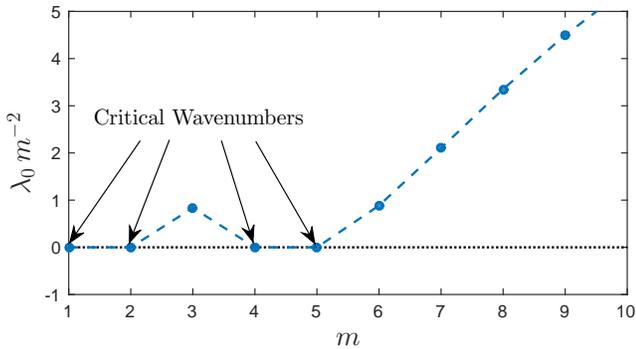}
\caption{\label{F:CriticalModes2D} (Color Online) Selected ground-state
 eigenvalues $\lambda_0$ of the linear operators generating the quadratic
 form $\mathcal{Q}_m$ at $\Gr=10^4$. The values of $\lambda_0$ correspond
 to the minimum of $\mathcal{Q}_m$ subject to the unit--norm constraint
 $\norm{W_m}^2_2=1$. %The critical modes are highlighted and the corresponding
% eigenfunctions are plotted in Figure~\ref{F:Minimisers2D}.
 }
\end{figure}

\begin{figure*}
\includegraphics[width=\textwidth, trim=4cm 2cm 4cm 0.5cm]{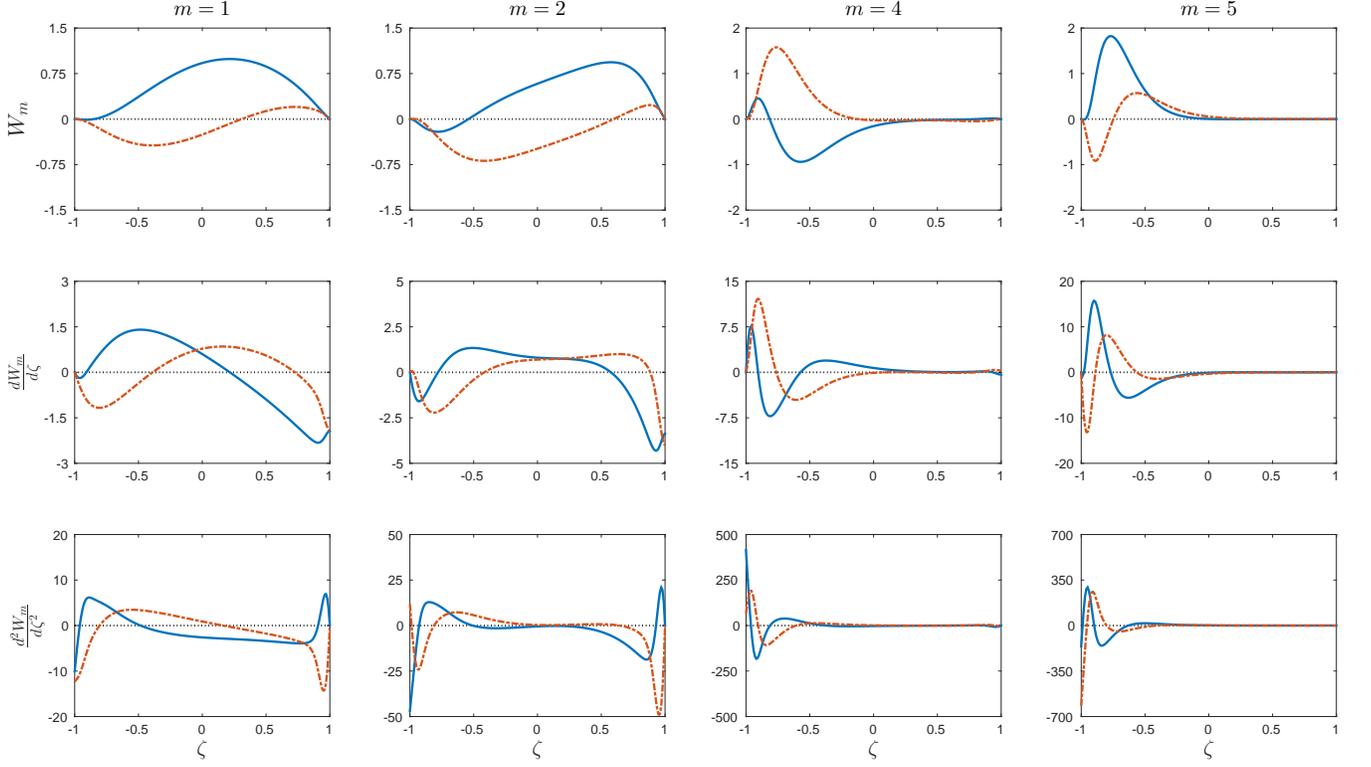}
\caption{\label{F:Minimisers2D} (Color Online) Real part (solid line) and
 imaginary part (dot-dashed line) of the critical ground state eigenfunctions
 and their derivatives at $\Gr=10^4$, normalized so that $\norm{W_m}^2_2=1$.}
\end{figure*}

The slight oscillations in the numerical bounds for 
$10^2\lesssim \Gr \lesssim 10^{4}$ are due to the occurrence of bifurcations
 in the number of critical Fourier modes, that is the number of values $m$
 such that the minimum of $\mathcal{Q}_m$ over non trivial functions is zero.
 This minimum coincides with the ground state eigenvalue $\lambda_0$ of the
 linear operator associated with $\mathcal{Q}_m$, and corresponds to the
 minimum eigenvalue of the matrix $\vec{Q}_m$, rescaled so that the respective
 eigenvector defines an eigenfunction $W_m$ with unit $L^2$ norm.
 Some selected values of $\lambda_0$ computed for $\Gr=10^4$ are plotted
 in Figure~\ref{F:CriticalModes2D}, while the critical ground state
 eigenfunctions are shown in Figure~\ref{F:Minimisers2D}. 
 Note that although conditions~\eqref{E:2DSufficientConditions} do not
 enforce~\eqref{E:AdmissibilityBCs_2} explicitly, the 
 critical modes satisfy all the correct BCs.

\subsection{Computational issues \label{SS:ComputIssues}}

As already mentioned, the range of Grashoff numbers we could study does not 
stretch into the expected asymptotic regime. This (current) limitation is a 
drawback of our method, as one would hope to be able to extract accurate scaling 
laws for large values of $\Gr$. 

A first limiting factor is that although the memory requirements for our largest 
SDP are modest and the computation times reasonable, we observed that the 
solution of the SDP became less accurate and prone to numerical ill conditioning 
as we increased $\Gr$ (and, consequently, the values of $P$, $N$ and $m_0$).
We expect that careful rescaling of the SDP data might help to resolve this issue;
yet, a general procedure is not available, and we leave the development
of an appropriate rescaling strategy to future work.

A second known issue is that the algorithms implemented in general purpose SDP 
solvers such as SeDuMi do not scale well with increasing problem size (see 
e.g.~\cite{Fukuda2000,Wen2010} for more details), and we expect that they will 
become unsuitable to solve the large SDPs needed at very large 
Grashoff numbers.
Addressing the challenges posed by large SDPs is the subject of a very active 
field of research (see 
e.g.~\cite{Fukuda2000,Nakata2003,Burer2003,
Burer2005,Burer2006,Wen2010,Sun2014}). In particular, the used of dedicated
algorithms instead of general purpose SDP solvers should be considered in 
future investigations, but is beyond the scope of the present work.

Finally, the task of validating the solution returned by the SDP solver might 
pose a challenge of its own if  memory requirements and computation time are a 
constraint. This is due to the cost of performing a large number 
of eigenvalue computations to check that the (generally large) matrices 
$\vec{Q}_m(\gvec{\hat{\phi}},\vec{t})$ are positive semidefinite for all $m$'s 
up to the correct $m_c$. For example, at $\Gr=10^5$ and with $P=34$, 
$N_\mathrm{checks}=350$, we had to compute the eigenvalues of a $774\times 774$ 
matrix 116 times for $\Gamma_x=2$ and 174 times for $\Gamma_x=3$ (cf. 
Table~\ref{T:Data2D}); in these cases, validating the solution was the most 
time-consuming task of the entire computation. In this case, a more careful 
estimate for $m_c$ might be helpful.

%%%%%%%%%%%%%%%%%%%%%%%% 
%    3D BOUNDS
%%%%%%%%%%%%%%%%%%%%%%%%
\section{Bounds for the three dimensional flow \label{S:3DBounds}}

To study the three dimensional flow, we follow~\cite{Tang2004,Hagstrom2014}
 and make the reasonable (although unproven) assumption that the critical
 modes determining the background field are independent of $x$ and,
 consequently, of the aspect ratio in the $x$ direction $\Gamma_x$.
 This assumption is not necessary for our method, but it simplifies the
 following analysis. Moreover, and most importantly, it reduces
 the size of the SDP we need to solve, as well as the cost of our
 \textit{a posteriori} checks. 
 This allows us to consider a wider range of $\Gr$
 at a reasonable computational cost, and therefore to draw relevant conclusions
 regarding the flow properties. 
 %As mentioned in Section~\ref{SS:ComputIssues}, addressing the 
 %computational issues associated with large SDPs is left for future work.

After setting $\alpha_m=0$ in Section~\ref{SS:SCFourier} and dropping the
 suffix $m$ to simplify the notation, the incompressibility condition allows
 us to eliminate $V_n$ and rewrite the quadratic forms $\mathcal{Q}_{n}$ as
\begin{align}
\mathcal{Q}_{n} = \int\limits_{-1}^{1} &\left[ \beta_n^2 \abs{U_{n}}^2 
+  4 \abs{\der{U_{n}}{\zeta}}^2 +  \beta_n^2\abs{W_{n}^2}  
+ 8 \abs{\der{W_{n}}{\zeta}}^2  \right.
\nonumber\\
&\left.   + \frac{16}{\beta_n^2} \abs{\der{^2W_{n}}{\zeta^2}}^2 
+ 4\der{\phi}{\zeta} \Re\left(U_{n} W_{n}^* \right) \right] d\zeta,
\label{E:SpectralConstraint3DFourier}
\end{align}
where the complex functions $U_n$ and $W_n$ satisfy the homogeneous BCs
\begin{subequations}
\label{E:3DBCs}
\begin{gather}
U_n(-1) = \der{U_n}{\zeta}\Big\vert_{1}= 0,\\
W_n(-1)= W_n(1)= \der{W_n}{\zeta}\Big\vert_{-1}= 
\der{^2W_n}{\zeta^2}\Big\vert_1 = 0.
\end{gather}
\end{subequations}

Since the real and imaginary parts of $U_n$ and $W_n$ give independent and
 formally identical contributions to $\mathcal{Q}_n$, to show that
 $\mathcal{Q}_n\geq 0$ it suffices to assume that $U_n$ and $W_n$  are
 real functions. Moreover, as in the two dimensional case, it is enough
 to consider strictly positive $n$'s up to the critical value
\begin{equation}
n_c(\phi)  := \left\lfloor \frac{\Gamma_y}{\pi} 
\sqrt{\frac{1}{2}\norm{\der{\phi}{\zeta}}_\infty}\right\rfloor.
\end{equation}
Consequently, the optimal bounds on $\Ce$ are determined by the solution of
 the variational problem
\begin{equation}
\label{E:OptProblem3D}
\begin{aligned}
&&\min_{\phi} \quad \mathcal{B}\{\phi\}& \\
\text{s.t.}  &&\mathcal{Q}_n\{U_n,W_n,\phi\} \geq 0,
& && 1\leq n \leq n_c(\phi), \\
&& \phi(-1)=0,& \\
&& \der{\phi}{\zeta}\Big\vert_1 = \frac{Gr}{2}.& 
\end{aligned}
\end{equation}

\subsection{An SDP for the optimal bounds \label{SS:SDP3D}}

As for the two dimensional flow, the variational 
problem~\eqref{E:OptProblem3D} can be
 recast as an SDP. In particular, we can use the parametrization of the
 background field and the linear objective function of
 Sections~\ref{S:BFparametrisation}--\ref{SS:2DLinearCostFunction}.
 Following analogous steps to Section~\ref{SS:LMIRelaxation2D}, we can expand
 $U_n$ and $W_n$ with appropriate Legendre series, consider $N$ coefficients
 explicitly and show that 
\begin{align}
\label{E:QnLowerBound}
\mathcal{Q}_n \geq &\,\,\gvec{\hat{\omega}}^T \vec{Q}_n(\gvec{\hat{\phi}})
  \gvec{\hat{\omega}}
%\nonumber\\
%&
+ 4\left( 1-\kappa_n\|\gvec{\hat{\phi}}\|_1 \right) \norm{\tilde{u}_1}_2^2 
\nonumber\\
&+ \frac{16}{\beta_n^2}\left( 1-\rho_n\|\gvec{\hat{\phi}}\|_1 \right)
\norm{\tilde{w}_2}_2^2.
\end{align}
Here, $\gvec{\hat{\omega}}$ is a vector containing the Legendre coefficients
 of $\der{U_n}{\zeta}$ and $\der{^2W_n}{\zeta^2}$ after imposing the BCs,
 while $\tilde{u}_1$ and $\tilde{w}_2$ are the corresponding remainder
 functions. Moreover, $\vec{Q}_n$ is a real, symmetric matrix whose entries
 are affine in $\gvec{\hat{\phi}}$ and $\|\gvec{\hat{\phi}}\|_1$, while
 $\kappa_n$ and $\rho_n$ are positive constants that decay with $N$.
The details are similar to those of Section~\ref{SS:LMIRelaxation2D}, and
 are omitted for brevity; we refer the interested reader to the more general
 discussion of~\cite{Fantuzzi2015a}.

As in Section~\ref{SS:SDP2D}, we can introduce a vector of slack variables
 $\vec{t}$ and add the $2P+2$ constraints 
${\hat{\phi}_j-t_j\leq 0}$, ${\hat{\phi}_j+t_j\geq0}$
to remove any absolute values from the right-hand side
 of~\eqref{E:QnLowerBound}. We can then replace each functional inequality
 $\mathcal{Q}_n\geq 0$ with the LMI 
$\vec{Q}_n(\gvec{\hat{\phi}},\vec{t})\succeq 0$ and two linear inequalities,
 and compute near-optimal bounds on $\Ce$ at each $\Gr$ after solving the SDP
\begin{equation}
\label{E:SDP3D}
\begin{aligned}
&\min_{\eta,\gvec{\hat{\phi}},\vec{t}} 
\qquad \frac{2}{Gr} \eta - 4\hat{\phi}_0
\\
&\begin{aligned}
\text{s.t.} \qquad\qquad \vec{S}(\gvec{\hat{\phi}},\eta) \succeq 0,&\\
\vec{Q}_n(\gvec{\hat{\phi}},\vec{t})\succeq 0,&  &&&&1\leq n \leq n_c,\\
%1-\kappa_n\sum_{p=0}^{P}t_p \geq 0,&  &&&&1\leq n \leq n_c, \\
1-\kappa_n \vec{1}^T\vec{t} \geq 0,&  &&&&1\leq n \leq n_c, \\
%1-\rho_n\sum_{p=0}^{P}t_p \geq 0,&  &&&&1\leq n \leq n_c, \\
1-\rho_n \vec{1}^T\vec{t} \geq 0,&  &&&&1\leq n \leq n_c, \\
\hat{\phi}_j -t_j\leq 0,&  &&&&0\leq j \leq P,\\
\hat{\phi}_j +t_j\geq 0,&  &&&&0\leq j \leq P, \\
%\sum_{p=0}^{P} \hat{\phi}_p = \frac{\Gr}{2}.
\vec{1}^T\gvec{\hat{\phi}} = \frac{\Gr}{2}.
\end{aligned}
\end{aligned}
\end{equation}
%

%%%%%%%%%%%
% FIGURE HERE FOR GOOD PLACEMENT IN PDF 
% BUT IT IS FOR THE NEXT SECTION
\begin{figure*}
\includegraphics[width=0.32\textwidth, trim=1cm 0cm 0cm 0.0cm]{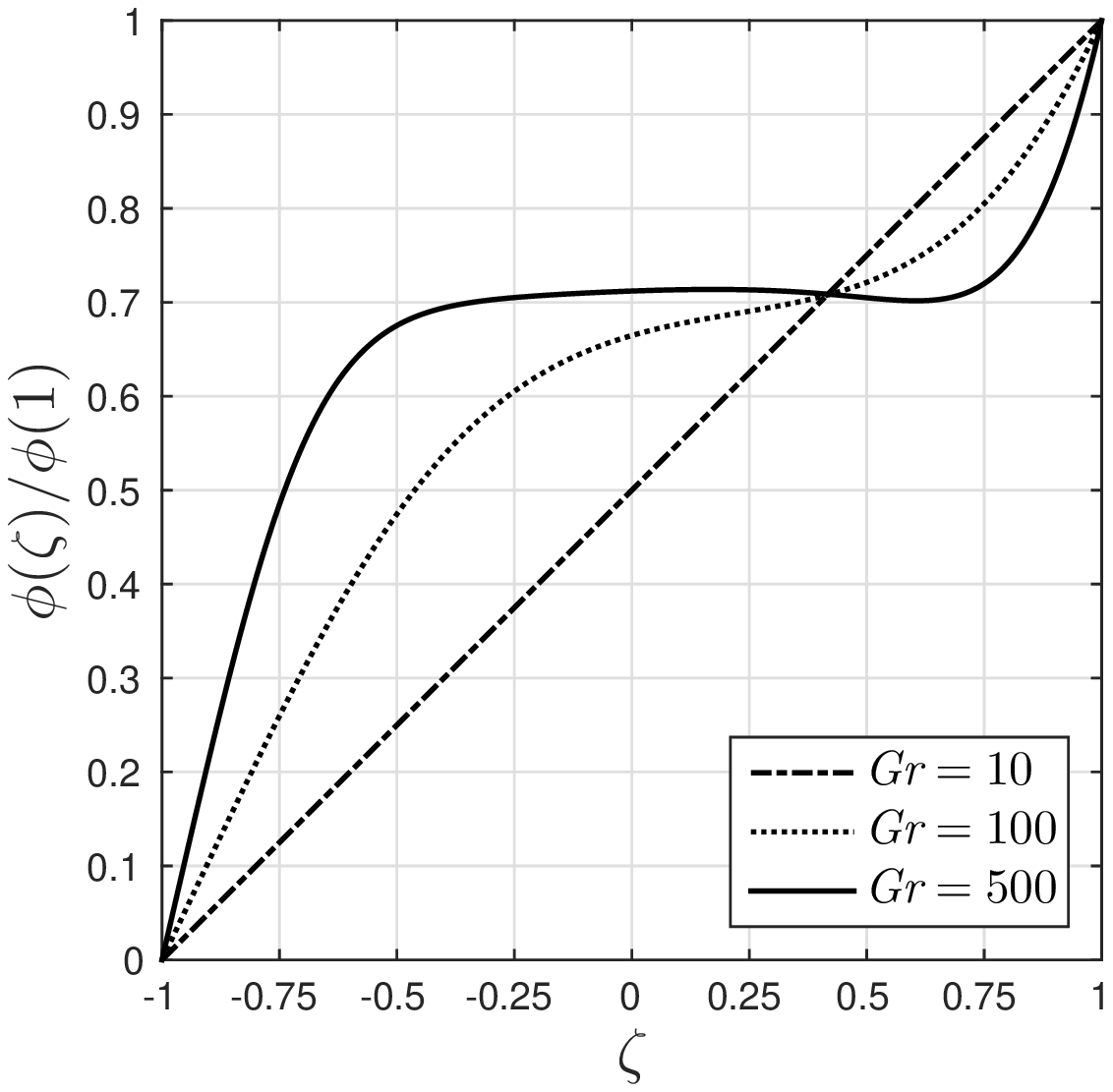}
%\hspace{15pt}
\includegraphics[width=0.32\textwidth, trim=0cm 0cm 1cm 0.0cm]{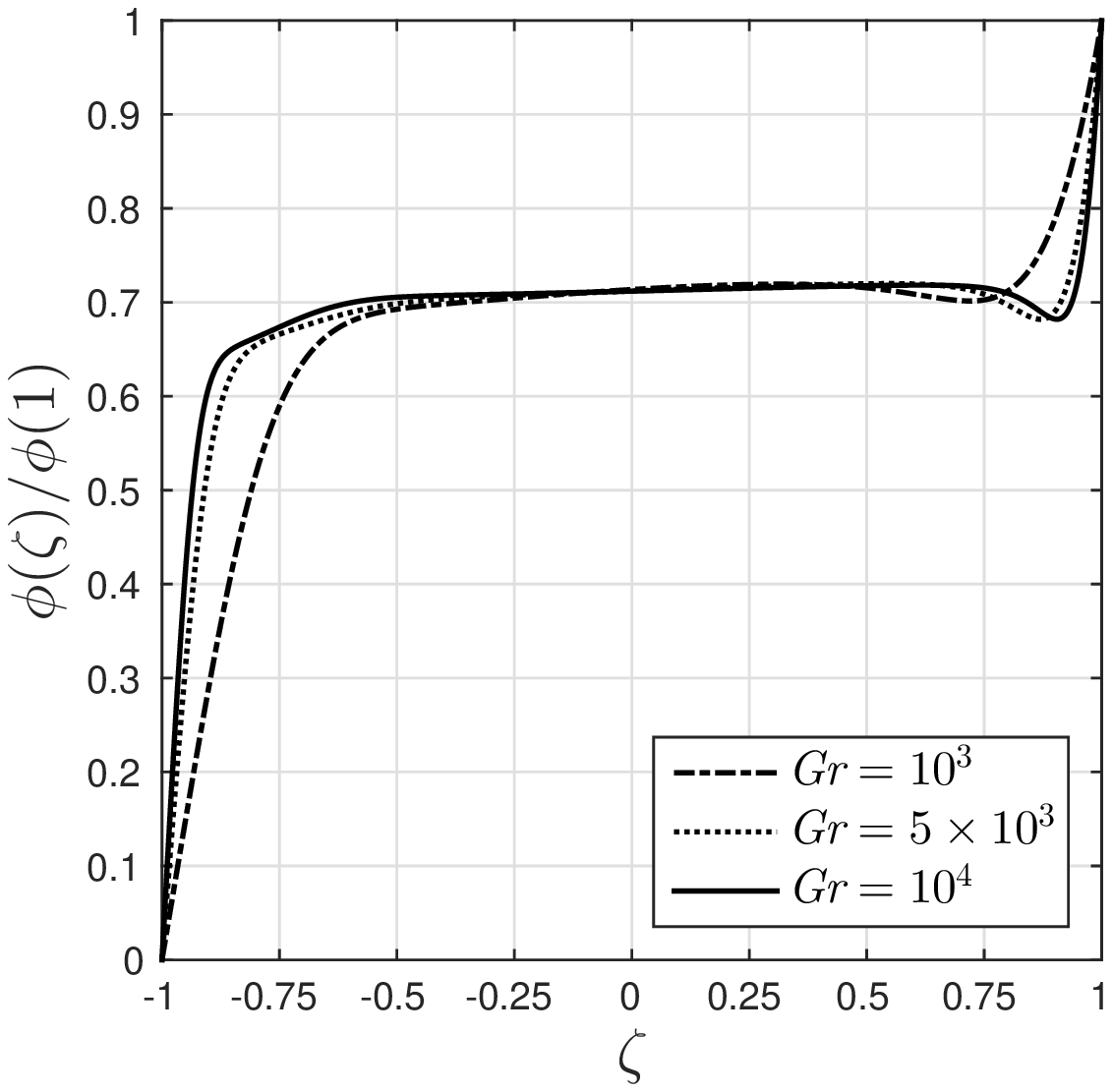}
\caption{\label{F:BP3DGammay3} Background fields normalized by their value
 at $\zeta=1$ for selected Grashoff numbers and $\Gamma_y=3$.}
\end{figure*}
%%%%%%%%%%%

As for the two dimensional case, we stress that the solution to this SDP gives
 a feasible background field for the infinite dimensional variational
 problem~\eqref{E:OptProblem3D} (modulo roundoff errors due to finite precision
 arithmetic).

Finally, as in Section~\ref{SS:SDP2D}, the number of inequalities $n_c$ is not 
known \emph{a priori}, so we will solve the optimization
using an initial guess $n_0$, verify the full set of LMIs 
\emph{a posteriori} as outlined in Section~\ref{SS:Results2D},
and repeat the optimization with an updated $n_c$ if such checks fail.

\subsection{Numerical implementation and results \label{SS:Results3D}}

\begin{table}[b]
  \caption{Problem data, memory requirements and CPU time for 
selected SDP instances. CPU time and memory include pre- and post-processing 
routines as well as the solution of the SDP.}
 \centering
 \begin{tabular}{ccccccccc}
 \hline\hline
 $\Gamma_y$ &  $\Gr$ & $n_0$ & $P$ & $N$ & RAM (Mb) &  Time (s) & $n_c$ & 
$N_{\mathrm{checks}}$\\
 \hline
 2 & $10^2$ &  5  & 34 & 100 &  15.1 &  55 & 5 & 100\\
 2 & $10^3$ & 10 & 39 & 125 &  46.9 & 168 & 14 & 125 \\
 2 & $10^4$ & 10 & 44 & 150 &  68.2 & 245 & 44 & 150\\
 \hline 
3 & $10^2$ &  5  & 34 & 100 &  15.1  & 53 & 7  & 100 \\
3 & $10^3$ & 10 & 39 & 125 &  46.9 & 105  & 21  & 125\\
3 & $10^4$ & 10 & 44 & 150 & 68.2  &  269 &   65 & 150 \\
\hline\hline
 \end{tabular}
\label{T:Data3D}
\end{table} 

The SDP~\eqref{E:SDP3D} was solved in MATLAB for $10\leq\Gr\leq 10^4$ and
 domain aspect ratios $\Gamma_y=2$, $\Gamma_y=3$. The problem specifications,
  memory requirements and computational time for 
  selected values of $\Gr$ are shown in 
Table~\ref{T:Data3D}. Note that, compared to the two dimensional case, 
higher polynomial degrees were required at a given $\Gr$
to resolve the finer structures characterizing the optimal background fields.
The details of the numerical implementation are as in 
Section~\ref{SS:Results2D}, and as in Section~\ref{SS:ComputIssues} 
the loss of accuracy in the solution of the SDP prevented us 
from reliably increasing $\Gr$ to larger values.

\begin{figure}[b]
\includegraphics[width=0.47\textwidth, trim=0cm 0.5cm 0cm 0cm]{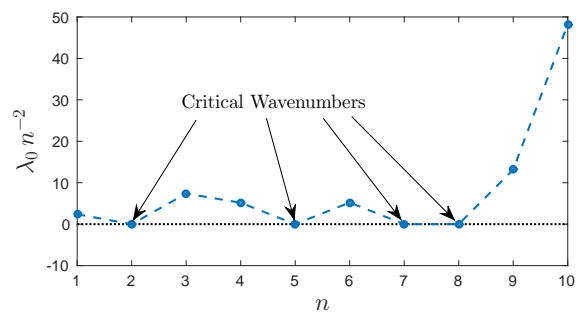}
\caption{\label{F:CriticalModes3D} (Color Online) Selected ground state
 eigenvalues $\lambda_0$ of the linear operators generating the quadratic
 form $\mathcal{Q}_n$ at $\Gr=10^4$. The values of $\lambda_0$ correspond
 to the minimum of $\mathcal{Q}_n$ subject to the unit norm constraint
 $\norm{U_n}^2_2+\norm{W_n}^2_2=1$. }
\end{figure}

A selection of optimal profiles for $\Gamma_y=3$ is shown in
 Figure~\ref{F:BP3DGammay3} (analogous results were obtained for
 $\Gamma_y=2$). Interestingly, as $\Gr$ is raised the boundary layer near the
 top boundary ($\zeta=1$) overshoots the approximately constant value in the
 bulk of the domain, resulting in a non-monotonic layer. 
 The boundary layer near $\zeta=-1$, instead, develops two regions
 with different characteristic slope (steeper near the boundary, 
 flatter towards the edge).
 We observed that the structural change leading to this ``internal layer" 
 near $\zeta=-1$ corresponds to the occurrence of bifurcations in the 
 number of critical Fourier modes in the SDP; 
 Figure~\ref{F:CriticalModes3D} shows that 3
 bifurcations have occurred at $\Gr=10^4$. A similar behavior 
 was also observed by Nicodemus \emph{et al.}~\cite{Nicodemus1998}, 
 who computed optimal
 background fields for the classical Couette flow, suggesting that the
 qualitative structural properties of the optimal background field at the
 bottom boundary are not affected by a change in the surface forcing.

\begin{figure*}
\includegraphics[width=0.75\textwidth, trim=0cm 0.5cm 0cm 0.5cm]{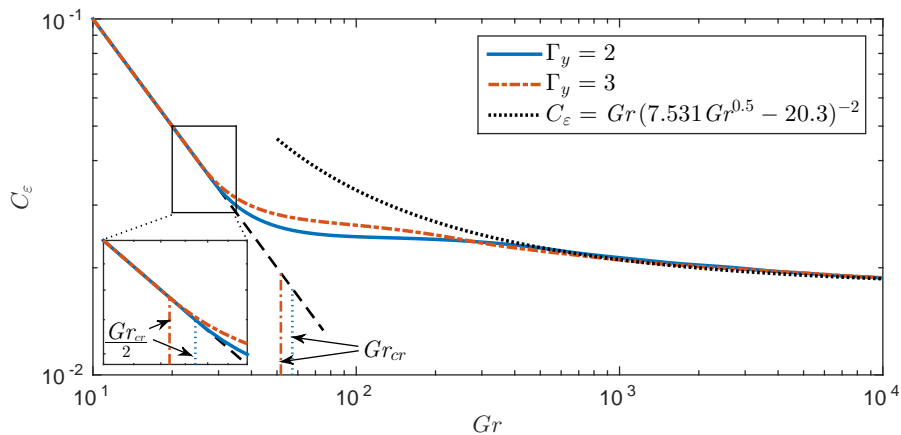}
\caption{\label{F:3DOptimalBounds} (Color Online) Optimal upper bounds on
 $\Ce$ for $\Gamma_y=2$ and $\Gamma_y=3$, compared to the numerical bounds
 from~\cite{Tang2004} (valid at large \Gr). The laminar dissipation
 coefficient is shown as a dashed black line, while the analytical bound
 $\Ce\lesssim  0.3536$ proven in~\cite{Hagstrom2014} is not plotted for
 clarity. Detail: as expected, our near-optimal bounds depart from the
 laminar value at $\Gr=0.5\Gr_{\text{cr}}$. }
\end{figure*}

The optimal bounds are plotted in Figure~\ref{F:3DOptimalBounds}, along with
 the laminar $\Ce$ and the asymptotic bounds estimated by Tang \emph{et al.}
 --- accurate for $\Gr\gtrsim 500$~\cite[Figure~3(b)]{Tang2004}. As for the
 two dimensional case, the bounds deviate from the laminar value at the
 expected value $\Gr = 0.5\Gr_{\text{cr}}$, where $\Gr_{\text{cr}}=57.20$ for
 $\Gamma_y=2$ and $\Gr_{\text{cr}}=51.73$ for $\Gamma_y=3$ 
 (cf. Section~\ref{S:EnergyStability}). This confirms that the solution of
 the SDP has converged to the optimal solution of~\eqref{E:OptProblem3D}.

The quantitative improvement compared to the analytical bound
 $\Ce\leq {1}/{(2\sqrt{2})}\approx  0.3536$ proven in~\cite{Hagstrom2014}
 (not plotted for clarity) is evident, our bounds being more than 10 times
 smaller at large $\Gr$. In addition, although the range of $\Gr$ we could
 study does not reach the asymptotic regime,
 it appears that for both values of $\Gamma_y$ the optimal bounds converge to
 the approximate results of~\cite{Tang2004}, which were computed by replacing
 the applied shear with a body force localized in a narrow region near the
 boundary. Hagstrom\,\&\,Doering demonstrated that this approximation
 does not change the energy stability boundaries of the laminar
 flow~\cite{Hagstrom2014}. Our results suggest that flows driven by a shear
 stress are similar to those driven by a body force in a narrow region near
 the upper surface also in terms of their energy dissipation rate --- at
 least when described by bounds on the dissipation coefficient. This is not
 surprising, since the background method, used to formulate the bounding
 problem for $\Ce$, can be seen as a generalization of energy stability theory.

%%%%%%%%%%%%%%%%%%%%%%%% 
%    ENERGY STABILITY
%%%%%%%%%%%%%%%%%%%%%%%%
\section{Semidefinite programming for energy 
stability problems \label{S:EnergyStability}}

The techniques used to compute the optimal background fields can also be
 applied directly to determine the energy stability boundaries of the  laminar
 Couette flow. The critical Grashoff number $\Gr_{\text{cr}}$ at which the
 solution is no longer energy stable is given by the maximization
 problem~\cite{Hagstrom2014}
\begin{equation}
\label{E:EnStabOptProblem}
\begin{gathered}
\max \quad Gr \\
\text{s.t.} \quad \stav{ \norm{\nabla\vec{u}}^2 + \Gr\, u w } \geq 0,
\end{gathered}
\end{equation}
where the spectral constraint is imposed over all horizontally-periodic, 
time-independent, incompressible velocity fields $\vec{u}(x,y,z)$ that satisfy
 the BCs in~\eqref{E:HomogBCs}.
The constraint can be relaxed as a combination of LMIs and linear inequalities
 using ides similar to those in Sections~\ref{S:2DBounds}--\ref{S:3DBounds}, 
and~\eqref{E:EnStabOptProblem} can be formulated as an SDP with one decision
 variable. The critical Grashoff $\Gr_{\text{cr}}$ can then be computed
 extremely efficiently, making semidefinite programming an attractive
 alternative to the traditional discretization of the boundary-eigenvalue
 problem associated with~\eqref{E:EnStabOptProblem}.

\begin{figure}[b]
\includegraphics[width=0.45\textwidth, trim=1.5cm 0cm 1.5cm 0cm]{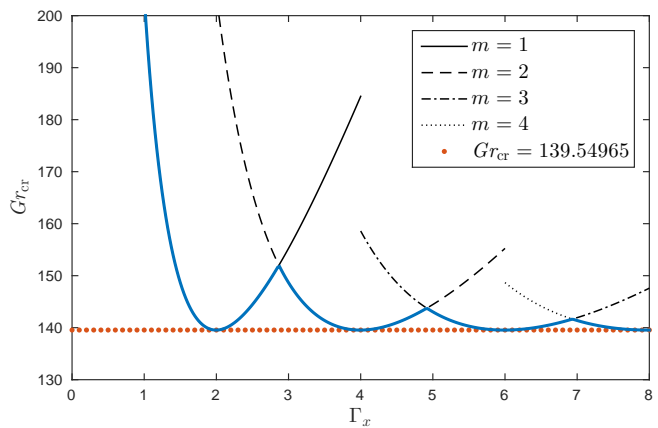}
\caption{\label{F:EnStab2D} (Color Online) Critical Grashoff numbers for
 energy stability of the Couette solution for the two dimensional flow.
 The neutral curves for individual Fourier modes and the results
 from~\cite{Hagstrom2014} are also shown for comparison.}
\end{figure}
\begin{figure}[t]
\includegraphics[width=0.45\textwidth, trim=1.5cm 0cm 1.5cm 0cm]{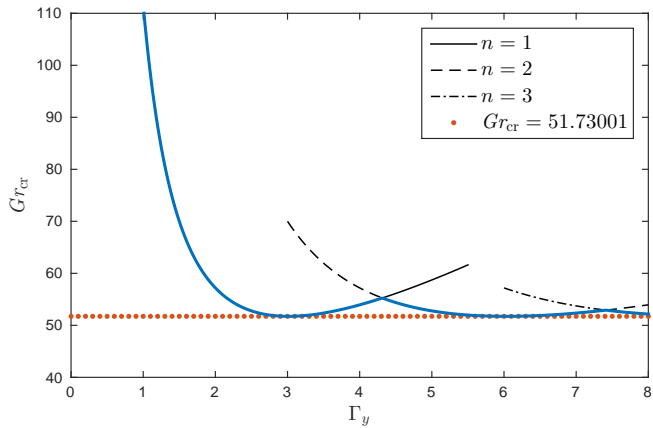}
\caption{\label{F:EnStab3D} (Color Online) Critical Grashoff numbers for
 energy stability of the Couette solution for the three dimensional flow,
 under the assumption that the critical mode is independent of the streamwise
 direction. The neutral curves for individual Fourier modes and the results
 from~\cite{Hagstrom2014} are also shown for comparison.}
\end{figure}

Figures~\ref{F:EnStab2D} and~\ref{F:EnStab3D} show $\Gr_{\text{cr}}$ for
 the two and three dimensional flows as a function of the
 domain aspect ratios $\Gamma_x$ and $\Gamma_y$. 
 As in Section~\ref{S:3DBounds}, we have assumed that the critical modes for
 the three dimensional flow are independent of the streamwise direction;
 this assumption is not necessary, but significantly simplifies the
 formulation of the SDP and reduces its computational cost. The figures
 also illustrate the neutral curves for each individual Fourier mode, which can be
 easily computed by considering only a single Fourier mode in the SDP. 
 Note that, as one would expect, all pairs $(m,\Gamma_x)$ achieving the same 
$\Gr_{\text{cr}}$ correspond to the same wavenumber $\alpha_m = 2\pi 
m/\Gamma_x$  (respectively, $(n,\Gamma_y)$ and $\beta_n = 2\pi n/\Gamma_y$ for 
the three dimensional flow). 
In particular, the local minima correspond to $\alpha_m 
\approx 3.15$ and $\beta_n \approx 2.1$ for the two and three dimensional flows 
respectively, in excellent agreement with the results of~\cite{Hagstrom2014}.

Strictly speaking, our results represent \textit{lower bounds} on
 $\Gr_{\text{cr}}$ that account for the effect of all orthogonal modes of 
 the velocity field, because the imposed LMIs are stronger conditions than the
 spectral constraint in~\eqref{E:EnStabOptProblem}. However, the optimal $\Gr$
 computed with our SDP formulation converges (from below) to the exact
 $\Gr_{\text{cr}}$ as the number of Legendre coefficients in the expansions
 of $\vec{u}$ is increased. In fact, we obtained well converged results
 considering as few as 15 Legendre coefficients. This is confirmed by the
 figures: our results agree extremely well with those obtained with
 traditional methods~\cite{Hagstrom2014} (these reference results were
 computed treating the Fourier wavenumber as a continuous variable and
 hence correspond to the limit of infinite aspect ratio).  %

%%%%%%%%%%%%%%%%%%%%%%%% 
%    CONCLUSIONS
%%%%%%%%%%%%%%%%%%%%%%%%
\section{Conclusions \label{S:Conclusions}}

Using a novel numerical technique, we have computed near-optimal bounds on
 the dissipation coefficient for two and three dimensional shear flows driven
 by a surface stress. To our knowledge, this is the first time that
 near-optimal bounds are determined for flows with imposed boundary fluxes;
 previous attempts have only considered piecewise linear fields
 (e.g.~\cite{Hagstrom2014,Wittenberg2010a}). 
Our numerical results improve previous analytical bounds by more than 10
 times at large $\Gr$, and agree with the approximate computations carried
 out by Tang~\textit{et al}~\cite{Tang2004}. This confirms that flows driven
  by a surface stress are similar to those driven
 by a localized body force not only in terms of their energy stability
 boundaries~\cite{Hagstrom2014}, but also as far as bounds on the
 dissipation coefficient are concerned. 

While our bounds have been obtained considering a restricted class of
 background fields, we expect that no significant further improvements can
 be achieved by other numerical techniques and, in practice, our results can
 be considered optimal. Yet, it should be recognized that our bounds are only
 optimal within the variational formulation proposed by
 Hagstrom\,\&\,Doering~\cite{Hagstrom2014}. In fact, as suggested by
 Tang~\textit{et al}.~\cite{Tang2004}, it is unlikely that our bounds are 
achieved by any real flow. It is therefore of future interest to compare our 
bounds to scaling laws estimated via numerical simulations or experiments of
 the flow at high $\Gr$ (not available at the time of writing).

Finally, we emphasize the central role played by our numerical approach, which
 allowed us to avoid the technical complications that affect the classical
 variational framework when fixed-flux BCs must be imposed. Although the
 computational challenges related to the solution of SDPs at large $\Gr$
 prevented us from reaching the
 asymptotic regime and estimating accurate scaling laws, the formulation
 of SDPs is generally attractive for the following two reasons. First, SDP
 relaxations can be derived systematically for a wide range of
 problems~\cite{Fantuzzi2015a}, not only of the type studied in this work.
 For instance, we have demonstrated that SDPs provide an efficient alternative
 to traditional methods, based on the solution of boundary-eigenvalue
 problems, in the context of energy stability.   
Second, the SDP is formulated in a rigorous manner, meaning that its
 feasible region corresponds to a set that is contained 
 within the feasible set of the original
 infinite dimensional bounding problem. While the bounds presented in this work
 cannot be considered analytic results due to the numerical roundoff errors in
 the solution of the SDPs, the rigorous formulation of a finite dimensional
 problem represents a first step towards the construction of a fully
 computer-assisted proof of near-optimal bounds.
For these reasons, we expect that if the technical difficulties related
 to solving SDPs in the asymptotic regime can be overcome by future work, 
 semidefinite programming will provide a robust framework to solve bounding 
 problems across a wide range of contexts.

%%%%%%%%%%%%%%%%%%%%%%%%%%%%%%%%%%%%%%%%%%%%%%%%%%%%%%%%%%%%%%%%%%%%%%%%%%%%%
%%%%%%%%%%%%%%%%%%%%%%%% 
%    APPENDIX
%\appendix
%\section{}
\appendix
\section{Properties of Legendre series \label{A:LegSeriesProperties} }

Let $W(\zeta)$ be defined for $\zeta\in[-1,1]$. We assume that $W$ has enough
 regularity such that $\der{W}{\zeta}$ has a uniformly convergent Legendre
 expansion; for example, assume that $W$ is twice continuously
 differentiable~\cite{Jackson1930}. Under the assumption of uniform
 convergence, the Legendre series
\begin{equation}
\begin{gathered}
W = \sum_{n=0}^{\infty} \hat{w}_n \leg_n(\zeta),
\\
\der{W}{\zeta} = \sum_{n=0}^{\infty} \hat{w}_n' \leg_n(\zeta),
\end{gathered}
\end{equation}
can be related using the fundamental theorem of calculus and the fact that
\begin{equation}
(2n+1) \leg_n(\zeta) = \der{}{\zeta} \left[ \leg_{n+1}(\zeta) 
- \leg_{n-1}(\zeta)\right]
\end{equation}
for all $n\geq 1$~\cite{Agarwal2009}. In fact, recalling that 
$\leg_0(\zeta)\equiv 1$ and $\leg_1(\zeta)\equiv \zeta$, we have
\begin{align}
W(\zeta) &= W(-1) + \int\limits_{-1}^{\zeta} \der{W}{t} dt 	\nonumber\\
&= W(-1) + \sum_{n=0}^{\infty} {\hat{w}'_n} \int\limits_{-1}^{\zeta} 
 \leg_{n}(t) dt 	\nonumber\\
&= W(-1)\cdot 1 + {\hat{w}'_0}(\zeta + 1) 
\nonumber\\
&\quad+ \sum_{n=1}^{\infty} \frac{\hat{w}'_n}{2n+1} \int\limits_{-1}^{\zeta}
  \der{}{\zeta} \left[ \leg_{n+1}(t) - \leg_{n-1}(t)\right] dt
\nonumber\\
&= W(-1)\leg_0(\zeta) + {\hat{w}'_0}\left[ \leg_1(\zeta) 
+ \leg_0(\zeta)\right] 
\nonumber\\
&\quad+ \sum_{n=1}^{\infty} \frac{\hat{w}'_n}{2n+1} \left[ \leg_{n+1}(\zeta) 
- \leg_{n-1}(\zeta)\right],
\end{align}
where the boundary values cancel out since $\leg_n(\pm1) = (\pm1)^n$. 
Rearranging the series and comparing coefficients we obtain 
the compatibility conditions
\begin{equation}
\begin{aligned}
\hat{w}_0 &= W(-1) + \hat{w}'_0 - \frac{\hat{w}'_1}{3}, \\
\hat{w}_n &= \frac{\hat{w}'_{n-1}}{2n-1} - \frac{\hat{w}'_{n+1}}{2n+3}, 
\quad n\geq 1.
\end{aligned}
\end{equation}
Moreover, it can be verified that
\begin{equation}
W(1) = W(-1) + \int\limits_{-1}^{1} \der{W}{\zeta} d\zeta= W(-1) 
+ 2\hat{w}'_0.
\end{equation}

It is clear that these expressions can be applied recursively to relate
 $W(\zeta)$ and the boundary value $W(1)$ to higher order derivatives under
 suitable regularity assumptions.

\section{Legendre expansion of the spectral constraint \label{A:LMIrelaxation}}

Substituting~\eqref{E:LegExp} into~\eqref{E:SpectralConstraint2DFourier}, 
recalling the definition of the remainder functions $\tilde{w}_0$,
 $\tilde{w}_1$, $\tilde{w}_2$, and using the orthogonality of the Legendre
 polynomials~\cite{Zeidler1995} we obtain
\begin{align}
\mathcal{Q}\{W,\phi\} &= 
\frac{16}{\alpha^2} \sum_{n=0}^{N+P+3}\frac{2\vert\hat{w}''_n \vert^2}{2n+1} 
+ 8 \sum_{n=0}^{N+1}\frac{2\vert\hat{w}'_n \vert^2}{2n+1} 
\nonumber\\
&\quad +
\alpha^2 \sum_{n=0}^{N}\frac{2\vert\hat{w}_n \vert^2}{2n+1}
- \Im\left( \mathcal{P}\right)
%\nonumber\\
%&\quad
+ \mathcal{R}.
\label{E:LMIrelaxationAppendixEqn1}
\end{align}
Here, $\mathcal{R}=\mathcal{R}\{\tilde{w}_0,\tilde{w}_1,\tilde{w}_2,\phi\}$ 
is as in~\eqref{E:Rdef} and
\begin{align}
\mathcal{P} := &\frac{8}{\alpha} \sum_{m=0}^{N+1} \sum_{n=0}^{\infty} 
\sum_{p=0}^{P} \hat{w}'_m \hat{w}_n^* \hat{\phi}_p \Lambda_{mnp}  
\nonumber\\ 
&+ \frac{8}{\alpha}\sum_{m=N+2}^{\infty} \sum_{n=0}^{N} \sum_{p=0}^{P}
 \hat{w}'_m \hat{w}_n^* \hat{\phi}_p \Lambda_{mnp},
\end{align}
where
\begin{equation}
\Lambda_{mnp} = \int\limits_{-1}^{1}\leg_m(\zeta) \leg_n(\zeta) 
\leg_p(\zeta)d\zeta.
\end{equation}
Applying the compatibility conditions~\eqref{E:IntegrationRule} to the first
 three terms of~\eqref{E:LMIrelaxationAppendixEqn1} we can find a real,
 symmetric, positive definite matrix $\vec{Q}_1$ such that
\begin{align}
\left(\vec{\hat{w}}''\right)^\dagger \vec{Q}_1 \vec{\hat{w}}'' =
 \,\,&\frac{16}{\alpha^2} \sum_{n=0}^{N+P+3}
\frac{2\vert\hat{w}''_n \vert^2}{2n+1} 
\nonumber\\
&+ 8 \sum_{n=0}^{N+1}\frac{2\vert\hat{w}'_n \vert^2}{2n+1} 
%\nonumber \\
+\alpha^2 \sum_{n=0}^{N}\frac{2\vert\hat{w}_n \vert^2}{2n+1}.
\end{align}
Moreover, since $\Lambda_{mnp}=0$ if $m-n+p<0$ or $n-m+p<0$~\cite{Dougall1953},
 the infinite sums over $n$ and $m$ in $\mathcal{P}$ can be truncated to
 $n\leq N+P+1$ and $m\leq N+P$ respectively. Calculating $\Lambda_{mnp}$ as
 in~\cite{Dougall1953} and letting
 \begin{equation}
 \gvec{\Phi}_{mn}(\gvec{\hat{\phi}}):= \frac{8}{\alpha}\sum_{p=0}^{P}
 \hat{\phi}_p\Lambda_{mnp}
 \end{equation} 
 we can write
\begin{align}
\mathcal{P} = 
&\sum_{m=0}^{N+1} \sum_{n=0}^{N+P+1}  \hat{w}'_m \hat{w}_n^* 
 \gvec{\Phi}_{mn} (\gvec{\hat{\phi}}) 
\nonumber \\
&+ \sum_{m=N+2}^{N+P} \sum_{n=0}^{N}  \hat{w}'_m \hat{w}_n^* 
\gvec{\Phi}_{mn}(\gvec{\hat{\phi}}).
\end{align}
It should be understood that the second term is zero if $P<2$.
Moreover, note that the $\gvec{\Phi}_{mn}$'s are linear in $\gvec{\hat{\phi}}$.
 Using~\eqref{E:IntegrationRule} we can then write
 $\mathcal{P} = \left(\vec{\hat{w}}''\right)^\dagger 
\vec{Q}_2(\gvec{\hat{\phi}}) \vec{\hat{w}}''$ for a suitably defined real matrix
 $\vec{Q}_2(\gvec{\hat{\phi}})$ whose entries are linear combinations
of the $\gvec{\Phi}_{mn}$'s (and are therefore linear in $\gvec{\hat{\phi}}$). 
We remark that, contrary to $\vec{Q}_1$, the matrix  
$\vec{Q}_2(\gvec{\hat{\phi}})$ is \emph{not} symmetric.

\section{A lower bound on $\mathcal{R}$ \label{A:LowerBoundR}}

In order to derive the lower bound on $\mathcal{R}$ stated
 in~\eqref{E:LowerBoundR}, we start by deriving a relation between
 $\norm{\tilde{w}_0}_2$, $\norm{\tilde{w}_1}_2$ and $\norm{\tilde{w}_2}_2$.
 Using~\eqref{E:IntegrationRule} and the elementary inequality
 $(a-b)^2\leq2a^2+2b^2$ we can write 
\begin{equation}
\begin{aligned}
\norm{\tilde{w}_1}_2^2 &= \sum_{n=N+2}^{\infty}
\frac{2\abs{\hat{w}_n'}^2}{2n+1} \\
&\leq \sum_{n=N+2}^{\infty}
\frac{4\abs{\hat{w}_{n-1}''}^2}{(2n-1)^2(2n+1)} \\
&\quad+ \sum_{n=N+2}^{\infty}
\frac{4\abs{\hat{w}_{n+1}''}^2}{(2n+1)(2n+3)^2}.
\end{aligned}
\end{equation}
Defining a matrix $\vec{H}_1$ such that
\begin{align}
(\vec{\hat{w}}'')^\dagger\vec{H}_1\vec{\hat{w}}'' &= 
\sum_{n=N+2}^{N+P+4}\frac{4\abs{\hat{w}_{n-1}''}^2}{(2n-1)^2(2n+1)}
 \nonumber\\
&\quad+ \sum_{n=N+2}^{N+P+2}
\frac{4\abs{\hat{w}_{n+1}''}^2}{(2n+1)(2n+3)^2}
\end{align}
and letting
\begin{equation}
\lambda_1 = \frac{4}{[2(N+P+4)-1][2(N+P+4)+3]}
\end{equation}
it can be verified that
\begin{equation}
\label{E:C4}
\norm{\tilde{w}_1}_2^2 \leq (\vec{\hat{w}}'')^\dagger\vec{H}_1\vec{\hat{w}}''
 + \lambda_1\norm{\tilde{w}_2}_2^2.
\end{equation}
Note that $\norm{\vec{H}_1}_F\sim\mathcal{O}(N^{-3})$ and
 $\lambda_1\sim\mathcal{O}(N^{-2})$.
Using similar ideas, we can express $\norm{\tilde{w}_0}_2^2$ in terms of
 the coefficients $\hat{w}_n'$, $n\leq N+1$ and $\norm{\tilde{w}_1}_2^2$.
 We then use~\eqref{E:IntegrationRule} and~\eqref{E:C4} to construct a
 matrix $\vec{H}_0$ and a constant $\lambda_0$ such that
\begin{equation}
\label{E:C5}
\norm{\tilde{w}_0}_2^2 \leq (\vec{\hat{w}}'')^\dagger\vec{H}_0\vec{\hat{w}}''
 + \lambda_0\norm{\tilde{w}_2}_2^2.
\end{equation}
Given the scaling estimates of $\vec{H}_1$ and $\lambda_1$ and the
 compatibility conditions~\eqref{E:IntegrationRule}, it is relatively
 straightforward to see that $\norm{\vec{H}_0}_F\sim\mathcal{O}(N^{-5})$
 and $\lambda_0\sim\mathcal{O}(N^{-4})$.

Let us now turn our attention to the functional $\mathcal{R}$ defined as
 in~\eqref{E:Rdef}. Clearly, we have the lower bound
\begin{equation}
\mathcal{R}\geq \frac{16}{\alpha^2}\norm{\tilde{w}_2}_2^2 
- \frac{8}{\alpha}\norm{\der{\phi}{\zeta}}_\infty
 \int\limits_{-1}^{1}\abs{\Im\left( \tilde{w}_1\tilde{w}_0^*\right)}d\zeta.
\end{equation}
Using Young's inequality, followed by~\eqref{E:C4} and~\eqref{E:C5}, we find
 that for any $\delta>0$

\begin{align}
\int\limits_{-1}^{1}\abs{\Im\left( \tilde{w}_1\tilde{w}_0^*\right)}d\zeta 
\leq &(\vec{\hat{w}}'')^\dagger \left[ \frac{\delta}{2}\vec{H}_0 
+ \frac{1}{2\delta}\vec{H}_1\right] \vec{\hat{w}}'' 
\nonumber\\
&+ \left[ \frac{\delta}{2}\lambda_0 + \frac{1}{2\delta}\lambda_1\right]
\norm{\tilde{w}_2}_2^2.
\end{align}
Letting
\begin{equation}
\delta = \sqrt{\frac{\lambda_1}{\lambda_0}},
\end{equation}
\begin{equation}
\vec{R} := \frac{8}{\alpha}\left[ \frac{\delta}{2}\vec{H}_0 
+ \frac{1}{2\delta}\vec{H}_1\right],
\end{equation}
\begin{equation}
\kappa = \frac{\alpha}{2}\sqrt{\lambda_0\lambda_1},
\end{equation}
and using~\eqref{E:InfNormBPEstimate} finally proves~\eqref{E:LowerBoundR}.

\section{Necessity of conditions~\eqref{E:2DSufficientConditions}
\label{A:CommentsSuffCond}}

Let us consider the problem of enforcing the constraint $\Psi\geq 0$, where 
$\Psi$ is as in equation~\eqref{E:PsiDefinition}, subject 
to~\eqref{E:AdmissibilityBCs}. Condition~\eqref{E:AdmissibilityBCs_1} can be 
enforced using~\eqref{E:BC1representation} as explained in 
Section~\ref{SS:LMIRelaxation2D}, and the problem reduces to showing that
\begin{align}
\label{E:Thesis}
\begin{bmatrix}
\Re\left(\gvec{\hat{\omega}}\right) \\  \Im\left(\gvec{\hat{\omega}}\right)
\end{bmatrix}^T
\vec{Q}(\gvec{\hat{\phi}})
\begin{bmatrix}
\Re\left(\gvec{\hat{\omega}}\right) \\  \Im\left(\gvec{\hat{\omega}}\right)
\end{bmatrix}
%\nonumber \\
%&
+\frac{16}{\alpha^2}\left( 1-\kappa \,\|\gvec{\hat{\phi}}\|_1\right)
 \norm{\tilde{w}_2}^2_2 \geq 0
\end{align}
for all $\gvec{\hat{\omega}} \in \R^{N+P+3}$ and $\tilde{w}_2(1)$ 
satisfying~\eqref{E:AdmissibilityBCs_2}. This condition can be rewritten with 
the help of~\eqref{E:BC1representation} as
\begin{equation}
\label{E:TransformedBC}
\tilde{w}_2(1) = - \vec{1}^T \vec{A}\gvec{\hat{\omega}},
\end{equation}
where $\vec{1}$ is a column vector of ones of length $N+P+4$. 

The conditions in~\eqref{E:2DSufficientConditions} are clearly sufficient 
for~\eqref{E:Thesis}, since they enforce the non-negativity of each of the two 
terms separately regardless of whether $\gvec{\hat{\omega}} $ and $\tilde{w}_2$ 
satisfy~\eqref{E:TransformedBC}.
Moreover, it is not difficult to see that~\eqref{E:SuffCondLinIneq} is also 
necessary. This follows because $\gvec{\hat{\omega}}=0$ is an admissible choice 
and there exists $\tilde{w}_2 \neq 0$ that satisfies $\tilde{w}_2(1)=0$.

To show that~\eqref{E:SuffCondLMI} is also necessary, for any fixed 
$\gvec{\hat{\omega}}$ we construct a sequence of remainder 
functions $\tilde{w}_2$ 
satisfying~\eqref{E:TransformedBC} showing that 
when~\eqref{E:Thesis} holds, the
term
\[
\begin{bmatrix}
\Re\left(\gvec{\hat{\omega}}\right) \\  \Im\left(\gvec{\hat{\omega}}\right)
\end{bmatrix}^T
\vec{Q}(\gvec{\hat{\phi}})
\begin{bmatrix}
\Re\left(\gvec{\hat{\omega}}\right) \\  \Im\left(\gvec{\hat{\omega}}\right)
\end{bmatrix}
\]
must be non-negative. In fact, given $\gvec{\hat{\omega}}$, 
consider the admissible remainder functions
\begin{equation}
\tilde{w}_2^{(m)}(\zeta) := \left( - \vec{1}^T \vec{A}\gvec{\hat{\omega}}
 \right)\mathcal{L}_m(\zeta)
\end{equation}
for $m\geq N+P+4$. For such a function, we have
\begin{align}
 \norm{\tilde{w}_2}^2_2 
 &= \frac{2 \vert \vec{1}^T \vec{A}\gvec{\hat{\omega}} \vert^2}{2m+1} 
 \nonumber \\
 &=\frac{2}{2m+1}  \gvec{\hat{\omega}} ^{\dagger} \left( \vec{A}^T \vec{1} 
\vec{1}^T \vec{A} \right) \gvec{\hat{\omega}} \nonumber\\
&\leq \lambda_{\mathrm{max}}\left( \vec{A}^T \vec{1} \vec{1}^T \vec{A} \right)
\norm{\gvec{\hat{\omega}}}^2,
\end{align}

where $\lambda_{\mathrm{max}}(\cdot)$ denotes the maximum eigenvalue of a 
matrix. Note that 
$\lambda_{\mathrm{max}}\left( \vec{A}^T \vec{1} \vec{1}^T \vec{A} \right)$
is a fixed positive quantity for any fixed $N$ and $P$.
If~\eqref{E:Thesis} holds, then
\begin{align}
\label{E:LBQeig}
\begin{bmatrix}
\Re\left(\gvec{\hat{\omega}}\right) \\  \Im\left(\gvec{\hat{\omega}}\right)
\end{bmatrix}^T
\vec{Q}(\gvec{\hat{\phi}})
\begin{bmatrix}
\Re\left(\gvec{\hat{\omega}}\right) \\  \Im\left(\gvec{\hat{\omega}}\right)
\end{bmatrix}
&\geq
-\frac{16}{\alpha^2}\left( 1-\kappa \,\|\gvec{\hat{\phi}}\|_1\right)
 \norm{\tilde{w}_2}^2_2 
 \nonumber \\
&\geq - \frac{C(\gvec{\hat{\phi}})}{2m+1} \norm{\gvec{\hat{\omega}}}^2,
\end{align}
where the non-negative term
\begin{equation}
C(\gvec{\hat{\phi}}) := \frac{32}{\alpha^2}
\lambda_{\mathrm{max}}\left( \vec{A}^T \vec{1} \vec{1}^T \vec{A} \right)
\left( 1-\kappa \,\|\gvec{\hat{\phi}}\|_1\right)
\end{equation}
only depends on the choice of $\gvec{\hat{\phi}}$, and can 
therefore be considered fixed.
Since~\eqref{E:LBQeig} holds for any $\gvec{\hat{\omega}}$, we must have
\begin{equation}
\lambda_{\mathrm{min}}\left[ \vec{Q}(\gvec{\hat{\phi}})  \right] \geq  -
\frac{C(\gvec{\hat{\phi}})}{2m+1},
\end{equation}
where $\lambda_{\mathrm{min}}(\cdot)$ denotes the minimum eigenvalue of a 
matrix. Letting $m\to \infty$ shows that $\vec{Q}(\gvec{\hat{\phi}})$ must be 
positive semidefinite, meaning that~\eqref{E:SuffCondLMI} is a necessary 
condition for~\eqref{E:Thesis} to hold for all $\gvec{\hat{\omega}}$'s 
and $\tilde{w}_2$'s subject to~\eqref{E:TransformedBC}.

%%%%%%%%%%%%%%%%%%%%%%%% 
%    ACKNOWLEDGEMENTS
\begin{acknowledgments}
The authors thank the anonymous reviewers, whose comments
 helped to improved the paper.
G.F. is grateful to Imperial College London for support under the
 IC PhD Scholarship, RESFS G82059.
\end{acknowledgments}

%%%%%%%%%%%%%%%%%%%%%%%%
%%%%%%%%%%%%%%%%%%%%%%%% 
%    REFERENCES
%\bibliography{references.bib}


\end{document}